\documentclass[10pt,aps,prl,a4paper,twocolumn,superscriptaddress]{revtex4-1}
  \usepackage[utf8]{inputenc}
\usepackage[colorlinks=true]{hyperref}
\usepackage{amsmath,amssymb}
\usepackage{amsfonts}
\usepackage{bbold}
\usepackage{graphicx}
\usepackage{color}
\usepackage[dvipsnames]{xcolor}
\usepackage{bm}
\usepackage[normalem]{ulem}

\date{\today}
\begin{document}

\title{Dynamics of bacteria scanning a porous environment}
\author{Ehsan Irani}
\affiliation{Max Delbrück Center for Molecular Medicine in the Helmholtz Association (MDC), The Berlin Institute for Medical Systems Biology (BIMSB), Berlin, Germany}
\author{Zahra Mokhtari}
\affiliation{Department of Mathematics and Computer Science, Freie Universit\"at Berlin}
\author{Annette Zippelius}
\affiliation{Institut f\"ur Theoretische Physik, Universit\"at G\"ottingen}
\date{\today}

\begin{abstract}
  It has recently been reported that bacteria, such as {\it
    E.coli}~\cite{datta2019} and {\it
    P. putida}~\cite{alirezaeizanjani2020chemotaxis}, perform distinct
  modes of motion when placed in porous media as compared to dilute
  regions or free space. This has led us to suggest an efficient
  strategy for active particles in a disordered environment:
  reorientations are suppressed in locally dilute regions and
  intensified in locally dense ones. Thereby the local geometry
  determines the optimal path of the active agent and substantially
  accelerates the dynamics for up to two orders of magnitude. We
  observe a non-monotonic behavior of the diffusion coefficient in
  dependence on the tumbling rate and identify a localisation
  transition, either by increasing the density of obstacles or by
  decreasing the reorientation rate.
\end{abstract}

\maketitle
The natural habitat of a wide range of microorganisms are complex
crowded media.  Examples are microorganisms which populate and
colonize rocks, modeled as micro-porous spaces, as well as bacteria
which contaminate or purify soil~\cite{dawid2000biology, wolfePNAS1989}.  In living matter, microorganisms
find themselves in a crowded environment, such as bacteria invading
mucus~\cite{cornick2015roles,laux2005role,cohen199524,celli2009helicobacter} or cells invading tissue~\cite{han2020cell}.  In many of these instances it is
vital for the microorganism to move efficiently through the porous and
tortuous environment, they are stuck in. The search for nutrients as
well as the escape from a poisonous environment has to be sufficiently
fast.  Many technical applications, such as water purification and
decomposition of contaminants trapped in the ground\cite{ginn2002processes,simon2002groundwater} rely on efficient dynamics of bacteria.  In medical applications bacteria are engineered
to sense the porous environment of a tumor~\cite{anderson2006environmentally,felfoul2016magneto} or microorganism are designed for drug delivery~\cite{luo2018micro}, -- fast and efficient dynamics of the bacteria being essential for their task.  Despite these widespread
applications, there is yet no consensus how the dynamics of such
organisms are adapted to perform most efficiently in a complex and
crowded medium. This has led us to ask: What is the best strategy for active agents to explore large porous regions in
short time? To what extent can adaptation to the inhomogeneous
environment accelerate the dynamics?

Several theoretical studies and simulations have addressed active
particles in a random environment \cite{Zoettl2019,zeitz2017active,reichhardt2014active,Bertrand2018,kurzthaler2021geometric,mokhtari2019dynamics}. Frequently the porous medium is modeled by the Lorentz
model~\cite{hofling2008critical,bauer2010localization}, where static
obstacles are placed randomly in space, covering a volume (area)
fraction $\phi_{\text{o}}$.
Zeitz et al.~\cite{zeitz2017active}
simulated active Brownian particles, whose
diffusion constant is depressed due to the tendency of active
particles to get stuck around obstacles.  Reichhardt et
al.~\cite{reichhardt2014active} 
include a drift term; surprisingly the drift velocity is non-monotonic
as a function of run time for given $\phi_{\text{o}}$. Bertrand et
al~\cite{Bertrand2018} compute the diffusion constant of active
particles in a lattice gas model and show that the diffusion constant
is non-monotonic in the tumbling rate as long as the obstacles are
static (or very slow). More recently Kurtzthaler et
al.~\cite{kurzthaler2021geometric} derived a geometric criterion for
optimal spreading, when the run length of the bacteria is comparable
to the longest straight path in the porous medium. In contrast to
these approaches, we suggest a {\bf local} adaptation mechanism of the
dynamics. Sensing the local density allows the microorganisms to
adjust their hopping rate efficiently in a strongly inhomogeneous
environment. Reorientation in dilute regions is ineffective and hence
suppressed; reorientation in dense regions and in particular in traps
is essential and hence fostered.

Local sensing of the environment has been adopted as a survival
mechanism in many phyla throughout the animal kingdom. Several
microorganisms regulate their behavior according to the density of
neighbours or to local gradients in phoretic propulsion.  For example,
a mechanism known as quorum sensing, allows bacteria to change their
speed according to the density of
neighbours~\cite{Liu2011,Fu2012,CatesPNAS2010,bauerle2018self,
  velasco2018collective, fischer2020quorum, rein2016collective,
  miller2001quorum}. Schools of fish have been observed to regulate their speed according to the density of neighbours~\cite{Katz2011,Mishra2012}.
 Regarding chemotaxis of bacteria in porous media, it has been suggested that the tumbling rate~\cite{Licata2016} as well as the tumbling angle~\cite{bhattacharjee2021chemotactic} changes in response to the local chemotactic concentration.
Bacteria with several swimming modes,
such as \emph{P. putida}~\citep{alirezaeizanjani2020chemotaxis}, can switch between different run modes in response to chemotactic conditions optimizing their chemotactic strategy. 
Volpe and Volpe~\cite{Volpe2017}
argue that the topography of the environment globally enhances the random
motion as compared to the ballistic one.
Recent experiments by Datta et al.~\cite{datta2019} on bacterial
hopping in porous media revealed that random disorder does not just
change the tumbling frequency and consequently also the run
length. Instead the bacteria are able to change their dynamics, if
trapped, so that hopping becomes dependent on the geometry of the pore
space.

{\it Model:}
We consider the dynamics of an elongated tracer particle in a two-dimensional medium
of static obstacles with area fraction $\phi_{\text{o}}$. The tracer is
modeled as a rigid trimer, consisting of 3 beads of radius $R_t$.  The
position vector of the central bead is denoted by $\bm{r}$. The two
peripheral beads are rigidly attached to the central bead, forming a
linear configuration, whose orientation is specified by a unit vector
$\bm{n}=(\cos{\varphi},\sin{\varphi})$. The position vectors of the two
peripheral beads are thus given by
$\bm{r}^{\pm}=\bm{r}\pm 2R_t \bm{n}$. The trimer is considered a model
for an elongated particle of aspect ratio 3.  The obstacles are
modeled as disks (2D), much larger than the beads of the trimer. In
the following we choose for the ratio of obstacle radius to tracer
radius $R_{\rm{o}}/R_t=10$. The interaction of the beads with the obstacles,
$\bm{F}({\bm{r}})$, is taken as a contact potential, modeled by a
stiff spring.

Since the trimer is modeled as a rigid body, its dynamics is fully
characterized by an equation for the translational motion of the center
of mass, which is taken to coincide with ${\bm{r}}$, and an
equation of motion for the orientation $\varphi$.
We assume over-damped dynamics, according to: 
\begin{equation}
\label{eq:overdamped}
\dot{\bm{r}} = \bm{v}_{\rm{a}}+\frac{1}{\gamma}
\sum_{i=1}^{N_{\rm{o}}}\bm{F}_i.
\end{equation}
The total force on the center of mass due to obstacle i at position
vector ${\bm{R}}_i$, is given
by
$\bm{F}_i=\bm{F}({\bm{r}}-{\bm{R}}_i)
+\bm{F}({\bm{r}^+}-{\bm{R}}_i)
+\bm{F}({\bm{r}^-}-{\bm{R}}_i)$.
The active velocity \(\bm{v}_{\rm{a}}\) is applied along the
direction of the trimer $\bm{n}$. Interactions with the obstacles cause the trimer to rotate:
\begin{equation}
\label{eq:orgc02e63e}
\dot\varphi = 
\sum_{i=1}^{N_{\rm{o}}}\tau_i.
\end{equation}
where the torque, $\bm{\tau}_i$, is explicitly given by
$\bm{\tau}_i=(\bm{r}^+-{\bm{r}})\times\bm{F}({\bm{r}^+}
-{\bm{R}}_i)+(\bm{r}^--{\bm{r}})\times\bm{F}({\bm{r}^-}-{\bm{R}}_i) $.
The torque is always normal to the plane of motion and $\tau_i$ is the projection of the vectorial torque on the normal of the plane of motion.

The occasional tumbling of bacteria has been modeled as a stochastic
reorientation process. For example, the bacteria reorient in random
directions with a given probability. Such a model is widely accepted
for run and tumble dynamics in solution. Does it apply also in dense
porous media? Recently it has been shown~\cite{datta2019} that bacterial dynamics
are changed when they are trapped. This has led us to introduce a
reorientation mechanism which depends on the local environment of the
tracer. In particular, the reorientations which disturb the
ballistic motion in void space and simultaneously prevent the particles from
getting trapped, are adapted to the local density in a strongly
heterogeneous environment. In that way, we try to model the
experimental finding that ``hops are guided by the geometry of the
pore space''~\cite{datta2019}.

Physical interactions between bacteria and surfaces are known to be determined by near-field lubrication forces~\cite{berke2008hydrodynamic, takagi2014hydrodynamic, sipos2015hydrodynamic} and steric collisions~\cite{drescher2011fluid}. Bacterial responses to such interactions vary from trapping in almost deterministic circular trajectories~\cite{takagi2014hydrodynamic} to enhanced reorientations~\cite{molaei2016succeed} depending on the type of surface and species. It has also been shown~\cite{fahrner2003bacterial, tipping2013load, wadhwa2019torque} that mechanical load on the flagella alters the flagellar motor kinematics, and thereby modulates reorientations.

\begin{figure}[!htpb]
\centering
\includegraphics[width=0.48\textwidth]{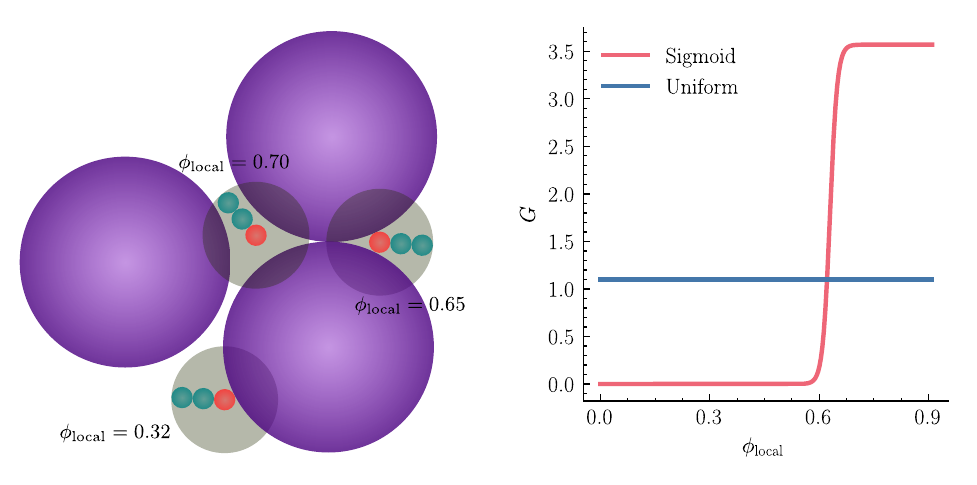}
\caption{\label{fig:orge764d0c}Left: A hypothetical disk (shown in grey) is
  defined for each trimer by a concentric area of radius \(r_l\)
  centered around the trimer's head. The local area fraction
  \(\phi_{\rm{local}}\) is the overlap of the hypothetical disk with
  neighbouring obstacles. Right: Tumbling rate $G$ for uniform
  (blue) and sigmoidal (red) $G$. }
\end{figure}

With a rate of $\lambda=1/ t_{\rm{re}}$ the local volume fraction \(\phi_{\rm{local}}\) is calculated inside a region
with radius \(r_{l}\), surrounding the trimer's head (see
Fig.\ref{fig:orge764d0c}). We use a function \(G (\phi_{\rm{local}})\)
to generate the probability of performing a random reorientation in the
full range from 0 to 2\(\pi\). The functional form of
\(G(\phi_{\rm{local}})\) encorporates the sensing mechanism which we refer to as ``density sensing`` in the following. Constant  $G$ results in the standard run and tumble dynamics with the rate of \( \lambda=1/ t_{\rm{re}} \), independent of the local environment. For a more sensitive function to $\phi_{\rm{local}}$ we consider a sigmoidal form as
  \begin{equation}
    G(\phi_{\rm{local}}) = \frac{C}{1+\exp^{-k(\phi_{\rm{local}} - \phi_{0})}}
    \label{sigmoid}
  \end{equation}
for the reorientation probability, with $C$ being the normalization factor. This choice reflects a high probability of reorientation in a locally dense region and a very low probability in a locally more dilute region. To approximate a step function, we choose $k=100$ and $\phi_0=0.63$ \\

We want to analyse the dynamics of the tracer particle as a function of several parameters. Without density dependent reorientation, these are
the magnitude of the active velocity \(v_{\rm{a}}\)
and the packing fraction of the obstacles \(\phi_{\text{o}}\)
Including density dependent reorientation, the important parameter is
the  time-scale \(t_{\rm{re}}\) of reorientation. The size of the region, $r_l$, to determine the local density should be comparable to  the size of the obstacle. Other functions \(G(\phi_{\text{o}})\), mapping \(\phi_{\rm{local}}\) to the probability of reorientation, may be considered in future work.

The parameters can be expressed in timescales. We measure lengths in units of $2R_{\rm{t}}$ and times in
units of the active timescale $t_a=2R_{\rm{t}}/v_{\rm{a}}$ which is controlled by the
active velocity. In these units, the reorientation time
$\tau_{\rm{re}}=t_{\rm{re}}/t_a$ is the Peclet number.  The collision
time is given $t_{\rm{coll}}^{-1}=2v_{\rm{a}}R_{\rm{o}}N_{\rm{o}}/L^2$ or in dimensionless units
$t_a/t_{\rm{coll}}=2R_{\rm{o}}R_{\rm{t}}N_{\rm{o}}/L^2$. It is controlled by the area fraction
$\phi_{\text{o}}=N_{\rm{o}}\pi R_{\rm{o}}^2/L^2$.
Both tumbling as well as collisions randomize the velocity of the
active particle and give rise to diffusion and hence cause a crossover
from ballistic motion to diffusive motion.

We used HOOMD-Blue~\cite{HOOMD2020} to integrate Eq.~\ref{eq:overdamped} and run Molecular Dynamics simulations on GPU ($\Delta t_{\rm{MD}} = 10^{-2}$). Freud package~\cite{freud2020} is used to investigate the local environment. For each set of ($\tau_{\rm{re}}$, $\phi_{\rm{o}}$), 10 to 50 simulations are performed, each with 100 independent trimers and $N_{\rm{o}}=2500$ random obstacles without overlaps. 

\begin{figure}[!htpb]
\centering
\includegraphics[width=0.45\textwidth]{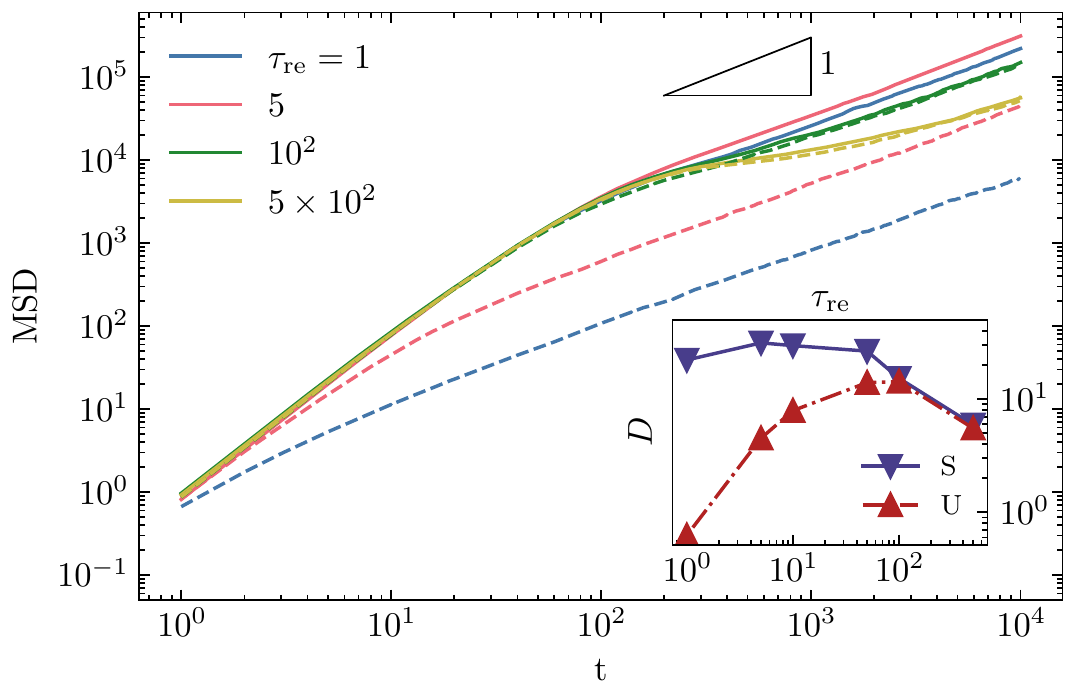}
\caption{\label{fig:msd-diff_t_re_diff_G} MSD of trimers for
  $\phi_{\text{o}}=0.4$ and different reorientation times $\tau_{\rm{re}}$; Comparison of
  sigmoidal (full line) and constant G (dashed line); inset: diffusion coefficient versus $\tau_{\rm{re}}$; a sigmoidal function $G$ (blue) is compared
  to a constant $G$ (red)}
\end{figure}

{\it Results:}
We focus here on the dynamics of tracer particles adapted to their
local environment.  In Fig.~\ref{fig:msd-diff_t_re_diff_G} we show the
MSD for $G(\phi_{\rm{local}})$ (full line) in comparison to a constant $G$
(dashed line). The most striking observation is the strong boost of
the dynamics for density sensing, when the reorientation time is
comparable to the timescale of active motion. The acceleration is due
to uninterrupted ballistic motion as well as reduced trapping
times. For moderate densities, such as $\phi_{\text{o}}=0.4$ shown in
Fig.~\ref{fig:msd-diff_t_re_diff_G}, the first mechanism dominates,
whereas for rather dense systems such as $\phi_{\text{o}}=0.7$, the latter
dominates (see below).

To quantify the acceleration due to density sensing, we extract a
diffusion constant as the slope of the MSD in the diffusive regime. It is
plotted in the inset of Fig.~\ref{fig:msd-diff_t_re_diff_G} as a function of $\tau_{\rm{re}}$ for both
uniform and sigmoidal $G$. The diffusion constant is larger by almost
two orders of magnitude for density sensing and $\tau_{\rm{re}}\sim 1$, i.e.
when the reorientation time is comparable to the timescale of active
motion. Furthermore, the diffusion constant is non-monotonic in
$\tau_{\rm{re}}$, as already observed in
Fig.~\ref{fig:msd-diff_t_re_diff_G}. The fastest dynamics is found for
$\tau_{\rm{re}}\sim 5$ and slows down for increasing as well as decreasing
$\tau_{\rm{re}}$. This non-monotonic behavior has been observed previously
for constant reorientation rate~\cite{Bertrand2018}, where it is in
fact more pronounced.  It can be explained by the following intuitive
argument: For large $\tau_{\rm{re}}$ the particles are stuck for a long
time in a locally dense region of obstacles, so that the diffusion
constant is small and approximately inversely proportional to
$\tau_{\rm{re}}$. For small $\tau_{\rm{re}}$, randomization of the active motion
is fast, so that the crossover from ballistic motion to diffusive
behavior happens at early times, resulting in small
values of the diffusion constant for small $\tau_{\rm{re}}$. These two
effects together give rise to an optimal value for $\tau_{\rm{re}}$, for
which the dynamics is fastest.

The difference between uniform and sigmoidal $G$ disappears for very
long reorientation times, when reorientation is so rare that ballistic
motion is mainly interrupted by collision events which are the same for
both models. 
\begin{figure}[!htpb]
\centering
\includegraphics[width=0.45\textwidth]{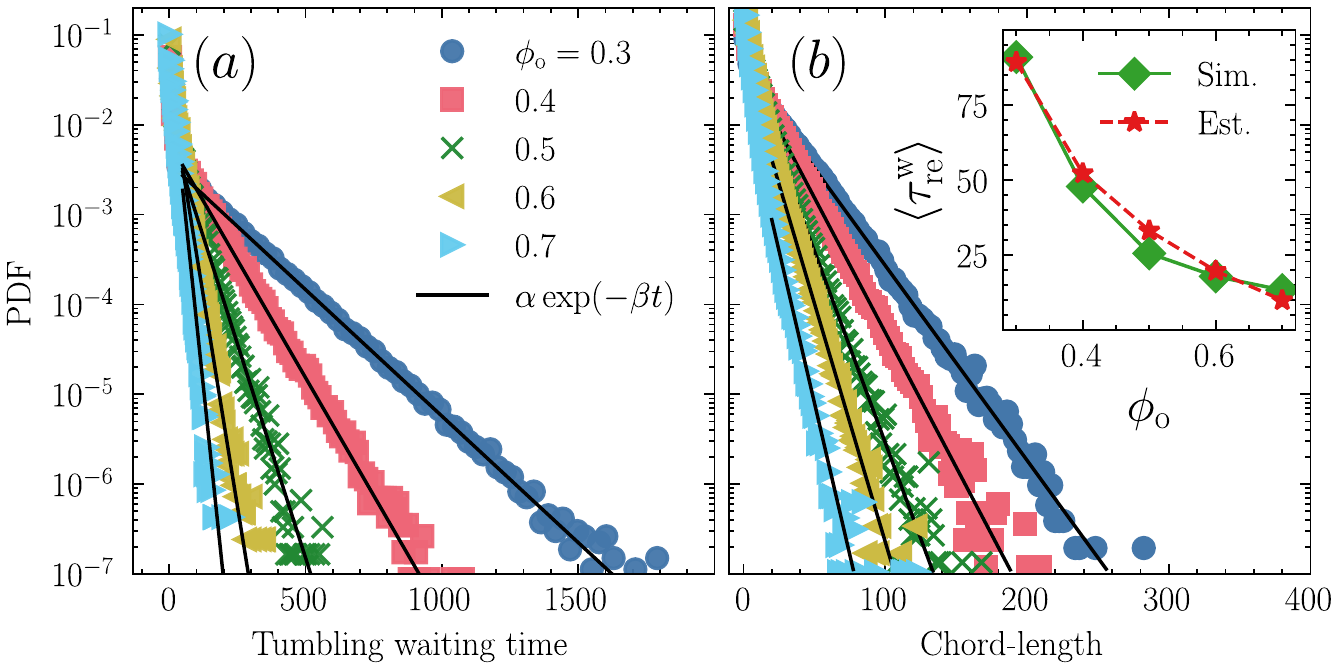}
\caption{\label{fig:hist_waiting_times} (a) Distribution of
  waiting times between two reorientation events. (b) Distribution of chord
  lengths; inset: Mean waiting time from the distribution (left) compared to estimate from chord lengths. All for
  different $\phi_{\text{o}}$, a sigmoidal $G$, and $\tau_{\rm{re}}=5$;}
\end{figure}
It has been suggested recently~\cite{kurzthaler2021geometric} that
reversing the velocity of the active particle is an efficient means to
accelerate the dynamics. For the sigmoidal $G$ random reorientation and run-reverse dynamics can hardly be distinguished (see Fig.1 in the SM)

The boost of the dynamics with density-sensing can be traced back to
the distribution of waiting times, defined as the time interval
between two reorientation events. The distribution is a simple
exponential for a uniform $G$ and all densities, characterized
uniquely by $\tau_{\rm{re}}$. In contrast, sigmoidal $G$ gives rise to
a second exponential, which slows down dramatically as the density
decreases, see Fig.~\ref{fig:hist_waiting_times}a.  In fact the decay
rate $\beta$ of the distribution increases approximately exponentially
with $\phi_{\text{o}}$.  The long relaxation times for
moderate $\phi_{\text{o}}$ are directly related to increasingly long
straight paths for dilute systems. Following
refs.\cite{Torquato1993,kurzthaler2021geometric} we compute the
distribution of straight paths which lie entirely in the void space of
the porous medium. The distribution of these so called chord lengths
is shown in Fig.~\ref{fig:hist_waiting_times}b. The distribution
strongly resembles the distribution of waiting times. 
  The average waiting time $\langle \tau_{\rm{re}}^w\rangle$ is the average distance the trimer travels from one trap to another, divided by the swimming speed. We estimate this distance as the mean chord length (extracted from Fig.~\ref{fig:hist_waiting_times}b) times the average number of collisions (evaluated by tracking trajectories) between two tumbling events. The inset of Fig.\ref{fig:hist_waiting_times}b compares this estimate for $\langle \tau_{\rm{re}}^w \rangle$ to the values computed from the distribution of waiting times (Fig.\ref{fig:hist_waiting_times}a).
The good agreement gives further support to our conclusion that the frequency of reorientation is determined by the geometry of the
environment. The latter determines the optimal path for the active
particle which is clearly seen in the movies in the
SM~\cite{suppmat}. Interestingly, the mean waiting time $\langle \tau_{\rm{re}}^w \rangle$ for optimal transport in our model ($ \tau_{\rm{re}} \sim 5$) lies in a similar range as the experimental data~\cite{datta2019} suggest (For details, see the SM~\cite{suppmat}).

\begin{figure}[!htpb]
\centering
\includegraphics[width=0.45\textwidth]{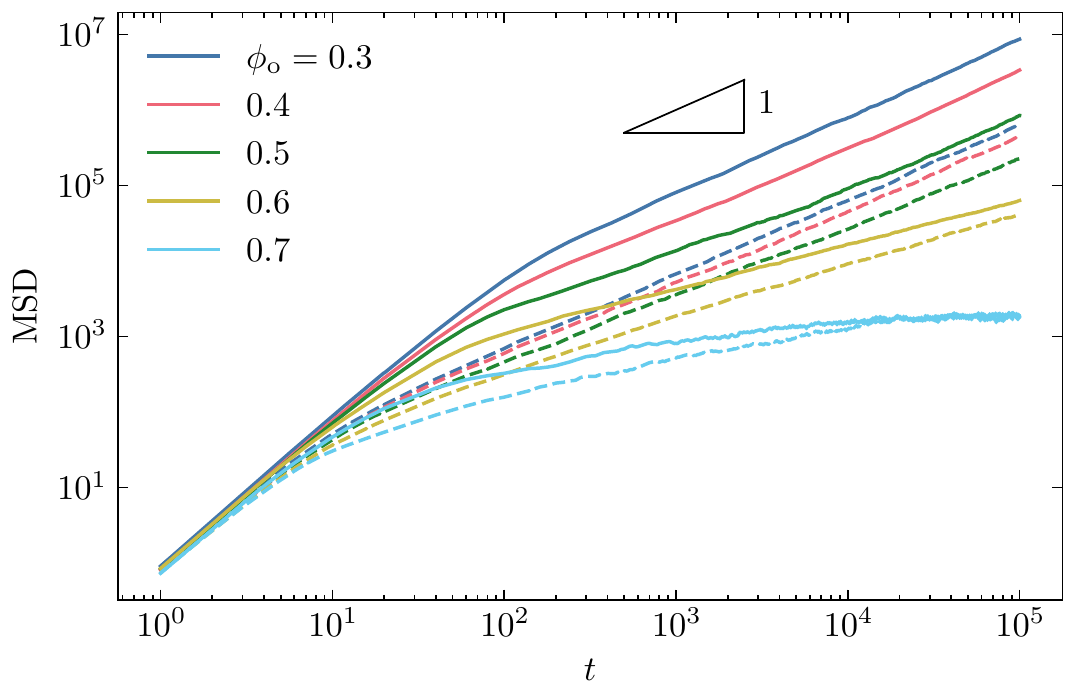}
\caption{\label{fig:msd-phi} MSD of trimers for
  $\tau_{\rm{re}}=5$ and different area fraction $\phi_{\text{o}}$; comparison of
  sigmoidal (full line) and constant G (dashed line). 
  }
\end{figure}

The dependence of the dynamics on the area fraction of obstacles is shown explicitly in
Fig.~\ref{fig:msd-phi}.
\begin{figure}[!hpbt]
\centering
\includegraphics[width=0.5\textwidth]{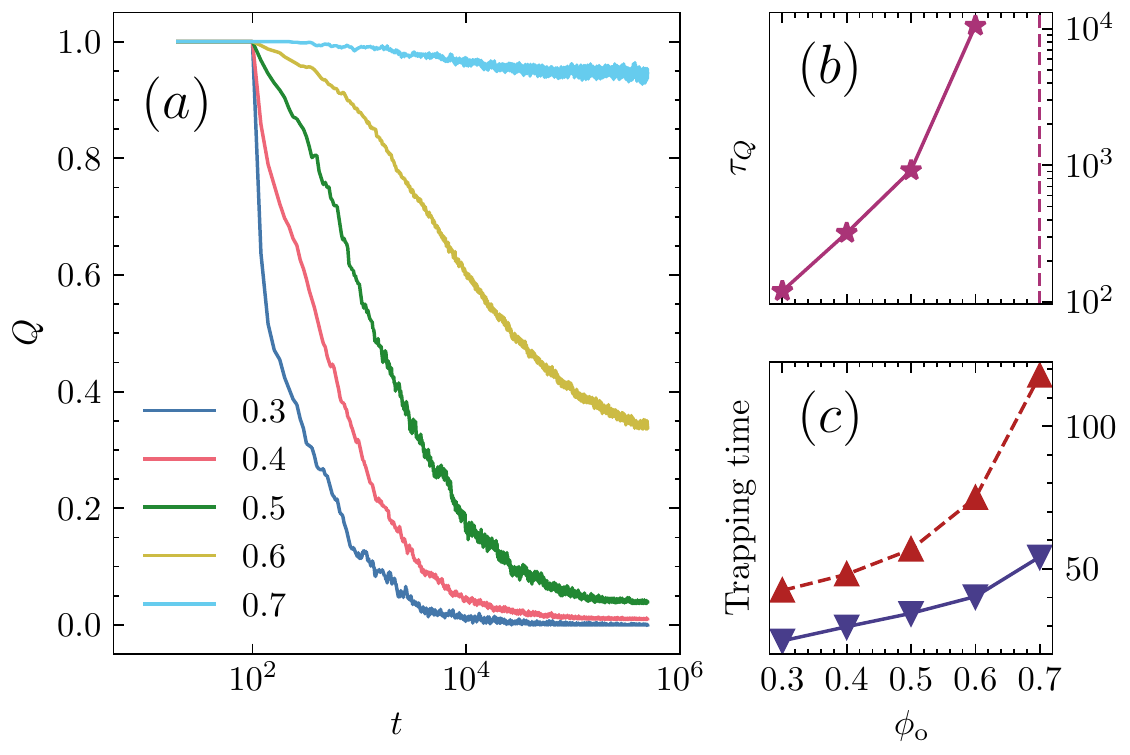}
\caption{\label{fig:Qtd-phi} (a) $Q(t;d)$ for $d=100$, $\tau_{\rm{re}}=5$ and several values of $\phi_{\text{o}}$ (color coding as in Fig.~\ref{fig:msd-phi}); (b)the relaxation time $\tau_Q$, defined by $Q(\tau_Q;d)=0.6$, versus $\phi_{\text{o}}$; (c) comparison of mean trapping time for uniform (red) and sigmoidal (blue) $G$ as a function of $\phi_{\text{o}}$.}
\end{figure}
As one expects, we see ballistic motion for short times and then a
crossover to diffusion at long times. As the density of obstacles
\(\phi_{\text{o}}\) is increased, the diffusion constant is reduced and
eventually at $\phi_{\text{o}}=0.7$, all particles are localised, exhibiting a
plateau in the MSD. The size of the plateau is the average squared
localisation length and independent of the dynamics. At the
localisation transition, we observe sub-diffusive behavior. The localisation transition in the Lorentz model with non-overlapping obstacles is a percolation transition. The
critical packing fraction for localisation, $\phi_{\rm{crit}}$, depends on the
size ratio $R_{\rm{o}}/R_{\rm{t}}$ and has been computed in~\cite{schnyder2015rounding}.

A finite fraction of particles is localised even for
$\phi_{\text{o}}<\phi_{\rm{crit}}$ due to the existence of finite size void
regions, coexisting with the macroscopic void area.  In fact the size
of these cages is widely distributed; for $\phi_{\text{o}}=0.4$ it
varies over more than two orders of magnitude (See Fig.~3 in SM~\cite{suppmat} for an example). A quantitative measure of partial localisation
is provided by $Q(t;d)$, the fraction of particles which have moved
less than $d$ in time $t$.  Choosing $d=0.3 L$, we observe that
$Q(t;d)$ decays to a finite value even well below
$\phi_{\text{o}}=0.7$, implying that a finite fraction of the
particles is localised (see Fig.~\ref{fig:Qtd-phi}a). As the
percolation transition is approached, the relaxation time of $Q(t;d)$
diverges, as can be seen in Fig.~\ref{fig:Qtd-phi}b), where we plot $\tau_Q$, defined as $Q(\tau_Q;d)=0.6$.
The fraction of localised particles, $Q_{\infty}(d)$ for $\phi_{\text{o}}\leq 0.7$ depends of course on the chosen value of
$d$. In the above figure this was chosen comparable to system size in
order to show that a finite fraction of particles is localised on
scales comparable to system size.

For dense systems, e.g. $\phi=0.6$, the dominant mechanism
responsible for the speed up of transport is the enhanced escape from cages.
Whereas for uniform $G$ particles reorient as much close to
the inner boundaries and dead ends of the cage as they do in the
center of the cage, for sigmoidal $G$ they effectively reorient only
at dead ends, resulting in a negligible number of reorientations,
unless it is necessary to escape a trap. Thereby, the mean trapping time is
reduced by a factor of approximately 2 for sigmoidal $G$ as compared
to uniform $G$ and most pronounced for the highest densities (see Fig.~\ref{fig:Qtd-phi}c)

Localisation can also occur for smaller $\phi_{\text{o}}$, such as
$\phi_{\text{o}}=0.4$, if the reorientation time is increased
accordingly. This is apparent in Fig.~\ref{fig:msd-diff_t_re_diff_G},
where we observe the emergence of a plateau for $\phi_{\text{o}}=0.4$
and $\tau_{\rm{re}}\geq 500$. Reorientations are a rare event, but
will take place for sufficiently long times.  Hence we expect to see a
crossover to diffusive behavior for even longer times, in contrast to
tracers above the percolation threshold (light blue curve in
Fig~\ref{fig:msd-phi}) which are truly localised.
Increasing $\tau_{\rm{re}}$ thus
provides another route to glassy dynamics in active matter (See Fig.~1 in SM~\cite{suppmat}).


We have introduced a model for bacterial spread in a porous medium,
which substantially accelerates the dynamics. It is based on a sensing
mechanism of the local density and thereby reduces adverse tumbling in
locally dilute regions and enhances necessary reorientations, when the
bacteria are trapped in local cages. 
The extremely long waiting times
between successive tumbling events for moderate densities can be
traced to the geometry of the porous structure which determines the
optimal path of the active agent.  For the fully random structure
under consideration, the diffusion constant can be enhanced by two
orders of magnitude. We expect the effect to be even stronger in a
structured system whose inhomogeneities extend over finite
length-scales.

The model can be easily extended to other transport phenomena which
require scanning of the environment. A prominent example is
chemotaxis, requiring local sensing of food or poison. Here a
concentration dependent tumbling rate may by the simplest model to
account for directed motion in a concentration gradient.

\begin{acknowledgments}
E.I acknowledges support from the Helmholtz Association (Germany), core funding to the Pombo group and computational resources at MDC. Z.M would like to acknowledge Germany’s Excellence Strategy – MATH+ : The Berlin Mathematics Research Center (EXC-2046/1) - project ID: 390685689 (subproject EF4-10) for partial support of this project.
\end{acknowledgments}

\bibliography{Bibliography}

\begin{thebibliography}{48}%
\makeatletter
\providecommand \@ifxundefined [1]{%
 \@ifx{#1\undefined}
}%
\providecommand \@ifnum [1]{%
 \ifnum #1\expandafter \@firstoftwo
 \else \expandafter \@secondoftwo
 \fi
}%
\providecommand \@ifx [1]{%
 \ifx #1\expandafter \@firstoftwo
 \else \expandafter \@secondoftwo
 \fi
}%
\providecommand \natexlab [1]{#1}%
\providecommand \enquote  [1]{``#1''}%
\providecommand \bibnamefont  [1]{#1}%
\providecommand \bibfnamefont [1]{#1}%
\providecommand \citenamefont [1]{#1}%
\providecommand \href@noop [0]{\@secondoftwo}%
\providecommand \href [0]{\begingroup \@sanitize@url \@href}%
\providecommand \@href[1]{\@@startlink{#1}\@@href}%
\providecommand \@@href[1]{\endgroup#1\@@endlink}%
\providecommand \@sanitize@url [0]{\catcode `\\12\catcode `\$12\catcode
  `\&12\catcode `\#12\catcode `\^12\catcode `\_12\catcode `\%12\relax}%
\providecommand \@@startlink[1]{}%
\providecommand \@@endlink[0]{}%
\providecommand \url  [0]{\begingroup\@sanitize@url \@url }%
\providecommand \@url [1]{\endgroup\@href {#1}{\urlprefix }}%
\providecommand \urlprefix  [0]{URL }%
\providecommand \Eprint [0]{\href }%
\providecommand \doibase [0]{http://dx.doi.org/}%
\providecommand \selectlanguage [0]{\@gobble}%
\providecommand \bibinfo  [0]{\@secondoftwo}%
\providecommand \bibfield  [0]{\@secondoftwo}%
\providecommand \translation [1]{[#1]}%
\providecommand \BibitemOpen [0]{}%
\providecommand \bibitemStop [0]{}%
\providecommand \bibitemNoStop [0]{.\EOS\space}%
\providecommand \EOS [0]{\spacefactor3000\relax}%
\providecommand \BibitemShut  [1]{\csname bibitem#1\endcsname}%
\let\auto@bib@innerbib\@empty
\bibitem [{\citenamefont {Bhattacharjee}\ and\ \citenamefont
  {Datta}(2019)}]{datta2019}%
  \BibitemOpen
  \bibfield  {author} {\bibinfo {author} {\bibfnamefont {T.}~\bibnamefont
  {Bhattacharjee}}\ and\ \bibinfo {author} {\bibfnamefont {S.}~\bibnamefont
  {Datta}},\ }\href@noop {} {\bibfield  {journal} {\bibinfo  {journal} {Nature
  Communications}\ }\textbf {\bibinfo {volume} {10}},\ \bibinfo {pages} {1}
  (\bibinfo {year} {2019})}\BibitemShut {NoStop}%
\bibitem [{\citenamefont {Alirezaeizanjani}\ \emph {et~al.}(2020)\citenamefont
  {Alirezaeizanjani}, \citenamefont {Gro{\ss}mann}, \citenamefont {Pfeifer},
  \citenamefont {Hintsche},\ and\ \citenamefont
  {Beta}}]{alirezaeizanjani2020chemotaxis}%
  \BibitemOpen
  \bibfield  {author} {\bibinfo {author} {\bibfnamefont {Z.}~\bibnamefont
  {Alirezaeizanjani}}, \bibinfo {author} {\bibfnamefont {R.}~\bibnamefont
  {Gro{\ss}mann}}, \bibinfo {author} {\bibfnamefont {V.}~\bibnamefont
  {Pfeifer}}, \bibinfo {author} {\bibfnamefont {M.}~\bibnamefont {Hintsche}}, \
  and\ \bibinfo {author} {\bibfnamefont {C.}~\bibnamefont {Beta}},\ }\href@noop
  {} {\bibfield  {journal} {\bibinfo  {journal} {Science advances}\ }\textbf
  {\bibinfo {volume} {6}},\ \bibinfo {pages} {eaaz6153} (\bibinfo {year}
  {2020})}\BibitemShut {NoStop}%
\bibitem [{\citenamefont {Dawid}(2000)}]{dawid2000biology}%
  \BibitemOpen
  \bibfield  {author} {\bibinfo {author} {\bibfnamefont {W.}~\bibnamefont
  {Dawid}},\ }\href@noop {} {\bibfield  {journal} {\bibinfo  {journal} {FEMS
  microbiology reviews}\ }\textbf {\bibinfo {volume} {24}},\ \bibinfo {pages}
  {403} (\bibinfo {year} {2000})}\BibitemShut {NoStop}%
\bibitem [{\citenamefont {Wolfe}\ and\ \citenamefont
  {Berg}(1989)}]{wolfePNAS1989}%
  \BibitemOpen
  \bibfield  {author} {\bibinfo {author} {\bibfnamefont {A.~J.}\ \bibnamefont
  {Wolfe}}\ and\ \bibinfo {author} {\bibfnamefont {H.~C.}\ \bibnamefont
  {Berg}},\ }\href {\doibase 10.1073/pnas.86.18.6973} {\bibfield  {journal}
  {\bibinfo  {journal} {Proceedings of the National Academy of Sciences}\
  }\textbf {\bibinfo {volume} {86}},\ \bibinfo {pages} {6973} (\bibinfo {year}
  {1989})}\BibitemShut {NoStop}%
\bibitem [{\citenamefont {Cornick}\ \emph {et~al.}(2015)\citenamefont
  {Cornick}, \citenamefont {Tawiah},\ and\ \citenamefont
  {Chadee}}]{cornick2015roles}%
  \BibitemOpen
  \bibfield  {author} {\bibinfo {author} {\bibfnamefont {S.}~\bibnamefont
  {Cornick}}, \bibinfo {author} {\bibfnamefont {A.}~\bibnamefont {Tawiah}}, \
  and\ \bibinfo {author} {\bibfnamefont {K.}~\bibnamefont {Chadee}},\
  }\href@noop {} {\bibfield  {journal} {\bibinfo  {journal} {Tissue barriers}\
  }\textbf {\bibinfo {volume} {3}},\ \bibinfo {pages} {e982426} (\bibinfo
  {year} {2015})}\BibitemShut {NoStop}%
\bibitem [{\citenamefont {Laux}\ \emph {et~al.}(2005)\citenamefont {Laux},
  \citenamefont {Cohen},\ and\ \citenamefont {Conway}}]{laux2005role}%
  \BibitemOpen
  \bibfield  {author} {\bibinfo {author} {\bibfnamefont {D.~C.}\ \bibnamefont
  {Laux}}, \bibinfo {author} {\bibfnamefont {P.~S.}\ \bibnamefont {Cohen}}, \
  and\ \bibinfo {author} {\bibfnamefont {T.}~\bibnamefont {Conway}},\
  }\href@noop {} {\bibfield  {journal} {\bibinfo  {journal} {Colonization of
  mucosal surfaces}\ ,\ \bibinfo {pages} {199}} (\bibinfo {year}
  {2005})}\BibitemShut {NoStop}%
\bibitem [{\citenamefont {Cohen}\ and\ \citenamefont
  {Laux}(1995)}]{cohen199524}%
  \BibitemOpen
  \bibfield  {author} {\bibinfo {author} {\bibfnamefont {P.~S.}\ \bibnamefont
  {Cohen}}\ and\ \bibinfo {author} {\bibfnamefont {D.~C.}\ \bibnamefont
  {Laux}},\ }in\ \href@noop {} {\emph {\bibinfo {booktitle} {Methods in
  enzymology}}},\ Vol.\ \bibinfo {volume} {253}\ (\bibinfo  {publisher}
  {Elsevier},\ \bibinfo {year} {1995})\ pp.\ \bibinfo {pages}
  {309--314}\BibitemShut {NoStop}%
\bibitem [{\citenamefont {Celli}\ \emph {et~al.}(2009)\citenamefont {Celli},
  \citenamefont {Turner}, \citenamefont {Afdhal}, \citenamefont {Keates},
  \citenamefont {Ghiran}, \citenamefont {Kelly}, \citenamefont {Ewoldt},
  \citenamefont {McKinley}, \citenamefont {So}, \citenamefont {Erramilli} \emph
  {et~al.}}]{celli2009helicobacter}%
  \BibitemOpen
  \bibfield  {author} {\bibinfo {author} {\bibfnamefont {J.~P.}\ \bibnamefont
  {Celli}}, \bibinfo {author} {\bibfnamefont {B.~S.}\ \bibnamefont {Turner}},
  \bibinfo {author} {\bibfnamefont {N.~H.}\ \bibnamefont {Afdhal}}, \bibinfo
  {author} {\bibfnamefont {S.}~\bibnamefont {Keates}}, \bibinfo {author}
  {\bibfnamefont {I.}~\bibnamefont {Ghiran}}, \bibinfo {author} {\bibfnamefont
  {C.~P.}\ \bibnamefont {Kelly}}, \bibinfo {author} {\bibfnamefont {R.~H.}\
  \bibnamefont {Ewoldt}}, \bibinfo {author} {\bibfnamefont {G.~H.}\
  \bibnamefont {McKinley}}, \bibinfo {author} {\bibfnamefont {P.}~\bibnamefont
  {So}}, \bibinfo {author} {\bibfnamefont {S.}~\bibnamefont {Erramilli}},
  \emph {et~al.},\ }\href@noop {} {\bibfield  {journal} {\bibinfo  {journal}
  {Proceedings of the National Academy of Sciences}\ }\textbf {\bibinfo
  {volume} {106}},\ \bibinfo {pages} {14321} (\bibinfo {year}
  {2009})}\BibitemShut {NoStop}%
\bibitem [{\citenamefont {Han}\ \emph {et~al.}(2020)\citenamefont {Han},
  \citenamefont {Pegoraro}, \citenamefont {Li}, \citenamefont {Li},
  \citenamefont {Yuan}, \citenamefont {Xu}, \citenamefont {Gu}, \citenamefont
  {Sun}, \citenamefont {Hao}, \citenamefont {Gupta} \emph
  {et~al.}}]{han2020cell}%
  \BibitemOpen
  \bibfield  {author} {\bibinfo {author} {\bibfnamefont {Y.~L.}\ \bibnamefont
  {Han}}, \bibinfo {author} {\bibfnamefont {A.~F.}\ \bibnamefont {Pegoraro}},
  \bibinfo {author} {\bibfnamefont {H.}~\bibnamefont {Li}}, \bibinfo {author}
  {\bibfnamefont {K.}~\bibnamefont {Li}}, \bibinfo {author} {\bibfnamefont
  {Y.}~\bibnamefont {Yuan}}, \bibinfo {author} {\bibfnamefont {G.}~\bibnamefont
  {Xu}}, \bibinfo {author} {\bibfnamefont {Z.}~\bibnamefont {Gu}}, \bibinfo
  {author} {\bibfnamefont {J.}~\bibnamefont {Sun}}, \bibinfo {author}
  {\bibfnamefont {Y.}~\bibnamefont {Hao}}, \bibinfo {author} {\bibfnamefont
  {S.~K.}\ \bibnamefont {Gupta}},  \emph {et~al.},\ }\href@noop {} {\bibfield
  {journal} {\bibinfo  {journal} {Nature physics}\ }\textbf {\bibinfo {volume}
  {16}},\ \bibinfo {pages} {101} (\bibinfo {year} {2020})}\BibitemShut
  {NoStop}%
\bibitem [{\citenamefont {Ginn}\ \emph {et~al.}(2002)\citenamefont {Ginn},
  \citenamefont {Wood}, \citenamefont {Nelson}, \citenamefont {Scheibe},
  \citenamefont {Murphy},\ and\ \citenamefont {Clement}}]{ginn2002processes}%
  \BibitemOpen
  \bibfield  {author} {\bibinfo {author} {\bibfnamefont {T.~R.}\ \bibnamefont
  {Ginn}}, \bibinfo {author} {\bibfnamefont {B.~D.}\ \bibnamefont {Wood}},
  \bibinfo {author} {\bibfnamefont {K.~E.}\ \bibnamefont {Nelson}}, \bibinfo
  {author} {\bibfnamefont {T.~D.}\ \bibnamefont {Scheibe}}, \bibinfo {author}
  {\bibfnamefont {E.~M.}\ \bibnamefont {Murphy}}, \ and\ \bibinfo {author}
  {\bibfnamefont {T.~P.}\ \bibnamefont {Clement}},\ }\href@noop {} {\bibfield
  {journal} {\bibinfo  {journal} {Advances in Water Resources}\ }\textbf
  {\bibinfo {volume} {25}},\ \bibinfo {pages} {1017} (\bibinfo {year}
  {2002})}\BibitemShut {NoStop}%
\bibitem [{\citenamefont {Simon}\ \emph {et~al.}(2002)\citenamefont {Simon},
  \citenamefont {Meggyes},\ and\ \citenamefont
  {T{\"u}nnermeier}}]{simon2002groundwater}%
  \BibitemOpen
  \bibfield  {author} {\bibinfo {author} {\bibfnamefont {F.-G.}\ \bibnamefont
  {Simon}}, \bibinfo {author} {\bibfnamefont {T.}~\bibnamefont {Meggyes}}, \
  and\ \bibinfo {author} {\bibfnamefont {T.}~\bibnamefont {T{\"u}nnermeier}},\
  }\href@noop {} {\bibfield  {journal} {\bibinfo  {journal} {Advanced
  groundwater remediation: active and passive technologies. Thomas Telford
  Publishing, London}\ ,\ \bibinfo {pages} {3}} (\bibinfo {year}
  {2002})}\BibitemShut {NoStop}%
\bibitem [{\citenamefont {Anderson}\ \emph {et~al.}(2006)\citenamefont
  {Anderson}, \citenamefont {Clarke}, \citenamefont {Arkin},\ and\
  \citenamefont {Voigt}}]{anderson2006environmentally}%
  \BibitemOpen
  \bibfield  {author} {\bibinfo {author} {\bibfnamefont {J.~C.}\ \bibnamefont
  {Anderson}}, \bibinfo {author} {\bibfnamefont {E.~J.}\ \bibnamefont
  {Clarke}}, \bibinfo {author} {\bibfnamefont {A.~P.}\ \bibnamefont {Arkin}}, \
  and\ \bibinfo {author} {\bibfnamefont {C.~A.}\ \bibnamefont {Voigt}},\
  }\href@noop {} {\bibfield  {journal} {\bibinfo  {journal} {Journal of
  molecular biology}\ }\textbf {\bibinfo {volume} {355}},\ \bibinfo {pages}
  {619} (\bibinfo {year} {2006})}\BibitemShut {NoStop}%
\bibitem [{\citenamefont {Felfoul}\ \emph {et~al.}(2016)\citenamefont
  {Felfoul}, \citenamefont {Mohammadi}, \citenamefont {Taherkhani},
  \citenamefont {De~Lanauze}, \citenamefont {Xu}, \citenamefont {Loghin},
  \citenamefont {Essa}, \citenamefont {Jancik}, \citenamefont {Houle},
  \citenamefont {Lafleur} \emph {et~al.}}]{felfoul2016magneto}%
  \BibitemOpen
  \bibfield  {author} {\bibinfo {author} {\bibfnamefont {O.}~\bibnamefont
  {Felfoul}}, \bibinfo {author} {\bibfnamefont {M.}~\bibnamefont {Mohammadi}},
  \bibinfo {author} {\bibfnamefont {S.}~\bibnamefont {Taherkhani}}, \bibinfo
  {author} {\bibfnamefont {D.}~\bibnamefont {De~Lanauze}}, \bibinfo {author}
  {\bibfnamefont {Y.~Z.}\ \bibnamefont {Xu}}, \bibinfo {author} {\bibfnamefont
  {D.}~\bibnamefont {Loghin}}, \bibinfo {author} {\bibfnamefont
  {S.}~\bibnamefont {Essa}}, \bibinfo {author} {\bibfnamefont {S.}~\bibnamefont
  {Jancik}}, \bibinfo {author} {\bibfnamefont {D.}~\bibnamefont {Houle}},
  \bibinfo {author} {\bibfnamefont {M.}~\bibnamefont {Lafleur}},  \emph
  {et~al.},\ }\href@noop {} {\bibfield  {journal} {\bibinfo  {journal} {Nature
  nanotechnology}\ }\textbf {\bibinfo {volume} {11}},\ \bibinfo {pages} {941}
  (\bibinfo {year} {2016})}\BibitemShut {NoStop}%
\bibitem [{\citenamefont {Luo}\ \emph {et~al.}(2018)\citenamefont {Luo},
  \citenamefont {Feng}, \citenamefont {Wang},\ and\ \citenamefont
  {Guan}}]{luo2018micro}%
  \BibitemOpen
  \bibfield  {author} {\bibinfo {author} {\bibfnamefont {M.}~\bibnamefont
  {Luo}}, \bibinfo {author} {\bibfnamefont {Y.}~\bibnamefont {Feng}}, \bibinfo
  {author} {\bibfnamefont {T.}~\bibnamefont {Wang}}, \ and\ \bibinfo {author}
  {\bibfnamefont {J.}~\bibnamefont {Guan}},\ }\href@noop {} {\bibfield
  {journal} {\bibinfo  {journal} {Advanced Functional Materials}\ }\textbf
  {\bibinfo {volume} {28}},\ \bibinfo {pages} {1706100} (\bibinfo {year}
  {2018})}\BibitemShut {NoStop}%
\bibitem [{\citenamefont {Z{\"o}ttl}\ and\ \citenamefont
  {Yeomans}(2019)}]{Zoettl2019}%
  \BibitemOpen
  \bibfield  {author} {\bibinfo {author} {\bibfnamefont {A.}~\bibnamefont
  {Z{\"o}ttl}}\ and\ \bibinfo {author} {\bibfnamefont {J.~M.}\ \bibnamefont
  {Yeomans}},\ }\href {\doibase 10.1038/s41567-019-0454-3} {\bibfield
  {journal} {\bibinfo  {journal} {Nature Physics}\ }\textbf {\bibinfo {volume}
  {15}},\ \bibinfo {pages} {554} (\bibinfo {year} {2019})}\BibitemShut
  {NoStop}%
\bibitem [{\citenamefont {Zeitz}\ \emph {et~al.}(2017)\citenamefont {Zeitz},
  \citenamefont {Wolff},\ and\ \citenamefont {Stark}}]{zeitz2017active}%
  \BibitemOpen
  \bibfield  {author} {\bibinfo {author} {\bibfnamefont {M.}~\bibnamefont
  {Zeitz}}, \bibinfo {author} {\bibfnamefont {K.}~\bibnamefont {Wolff}}, \ and\
  \bibinfo {author} {\bibfnamefont {H.}~\bibnamefont {Stark}},\ }\href@noop {}
  {\bibfield  {journal} {\bibinfo  {journal} {The European Physical Journal E}\
  }\textbf {\bibinfo {volume} {40}},\ \bibinfo {pages} {23} (\bibinfo {year}
  {2017})}\BibitemShut {NoStop}%
\bibitem [{\citenamefont {Reichhardt}\ and\ \citenamefont
  {Reichhardt}(2014)}]{reichhardt2014active}%
  \BibitemOpen
  \bibfield  {author} {\bibinfo {author} {\bibfnamefont {C.}~\bibnamefont
  {Reichhardt}}\ and\ \bibinfo {author} {\bibfnamefont {C.~O.}\ \bibnamefont
  {Reichhardt}},\ }\href@noop {} {\bibfield  {journal} {\bibinfo  {journal}
  {Physical Review E}\ }\textbf {\bibinfo {volume} {90}},\ \bibinfo {pages}
  {012701} (\bibinfo {year} {2014})}\BibitemShut {NoStop}%
\bibitem [{\citenamefont {Bertrand}\ \emph {et~al.}(2018)\citenamefont
  {Bertrand}, \citenamefont {Zhao}, \citenamefont {Benichou}, \citenamefont
  {Tailleur},\ and\ \citenamefont {Voiturez}}]{Bertrand2018}%
  \BibitemOpen
  \bibfield  {author} {\bibinfo {author} {\bibfnamefont {T.}~\bibnamefont
  {Bertrand}}, \bibinfo {author} {\bibfnamefont {Y.}~\bibnamefont {Zhao}},
  \bibinfo {author} {\bibfnamefont {O.}~\bibnamefont {Benichou}}, \bibinfo
  {author} {\bibfnamefont {J.}~\bibnamefont {Tailleur}}, \ and\ \bibinfo
  {author} {\bibfnamefont {R.}~\bibnamefont {Voiturez}},\ }\href@noop {}
  {\bibfield  {journal} {\bibinfo  {journal} {Phys. Rev. Lett.}\ }\textbf
  {\bibinfo {volume} {120}},\ \bibinfo {pages} {198103} (\bibinfo {year}
  {2018})}\BibitemShut {NoStop}%
\bibitem [{\citenamefont {Kurzthaler}\ \emph {et~al.}(2021)\citenamefont
  {Kurzthaler}, \citenamefont {Mandal}, \citenamefont {Bhattacharjee},
  \citenamefont {L{\"o}wen}, \citenamefont {Datta},\ and\ \citenamefont
  {Stone}}]{kurzthaler2021geometric}%
  \BibitemOpen
  \bibfield  {author} {\bibinfo {author} {\bibfnamefont {C.}~\bibnamefont
  {Kurzthaler}}, \bibinfo {author} {\bibfnamefont {S.}~\bibnamefont {Mandal}},
  \bibinfo {author} {\bibfnamefont {T.}~\bibnamefont {Bhattacharjee}}, \bibinfo
  {author} {\bibfnamefont {H.}~\bibnamefont {L{\"o}wen}}, \bibinfo {author}
  {\bibfnamefont {S.~S.}\ \bibnamefont {Datta}}, \ and\ \bibinfo {author}
  {\bibfnamefont {H.~A.}\ \bibnamefont {Stone}},\ }\href {\doibase
  10.1038/s41467-021-26942-0} {\bibfield  {journal} {\bibinfo  {journal}
  {Nature Communications}\ }\textbf {\bibinfo {volume} {12}},\ \bibinfo {pages}
  {7088} (\bibinfo {year} {2021})}\BibitemShut {NoStop}%
\bibitem [{\citenamefont {Mokhtari}\ and\ \citenamefont
  {Zippelius}(2019)}]{mokhtari2019dynamics}%
  \BibitemOpen
  \bibfield  {author} {\bibinfo {author} {\bibfnamefont {Z.}~\bibnamefont
  {Mokhtari}}\ and\ \bibinfo {author} {\bibfnamefont {A.}~\bibnamefont
  {Zippelius}},\ }\href@noop {} {\bibfield  {journal} {\bibinfo  {journal}
  {Physical review letters}\ }\textbf {\bibinfo {volume} {123}},\ \bibinfo
  {pages} {028001} (\bibinfo {year} {2019})}\BibitemShut {NoStop}%
\bibitem [{\citenamefont {H{\"o}fling}\ \emph {et~al.}(2008)\citenamefont
  {H{\"o}fling}, \citenamefont {Munk}, \citenamefont {Frey},\ and\
  \citenamefont {Franosch}}]{hofling2008critical}%
  \BibitemOpen
  \bibfield  {author} {\bibinfo {author} {\bibfnamefont {F.}~\bibnamefont
  {H{\"o}fling}}, \bibinfo {author} {\bibfnamefont {T.}~\bibnamefont {Munk}},
  \bibinfo {author} {\bibfnamefont {E.}~\bibnamefont {Frey}}, \ and\ \bibinfo
  {author} {\bibfnamefont {T.}~\bibnamefont {Franosch}},\ }\href@noop {}
  {\bibfield  {journal} {\bibinfo  {journal} {The Journal of chemical physics}\
  }\textbf {\bibinfo {volume} {128}},\ \bibinfo {pages} {164517} (\bibinfo
  {year} {2008})}\BibitemShut {NoStop}%
\bibitem [{\citenamefont {Bauer}\ \emph {et~al.}(2010)\citenamefont {Bauer},
  \citenamefont {H{\"o}fling}, \citenamefont {Munk}, \citenamefont {Frey},\
  and\ \citenamefont {Franosch}}]{bauer2010localization}%
  \BibitemOpen
  \bibfield  {author} {\bibinfo {author} {\bibfnamefont {T.}~\bibnamefont
  {Bauer}}, \bibinfo {author} {\bibfnamefont {F.}~\bibnamefont {H{\"o}fling}},
  \bibinfo {author} {\bibfnamefont {T.}~\bibnamefont {Munk}}, \bibinfo {author}
  {\bibfnamefont {E.}~\bibnamefont {Frey}}, \ and\ \bibinfo {author}
  {\bibfnamefont {T.}~\bibnamefont {Franosch}},\ }\href@noop {} {\bibfield
  {journal} {\bibinfo  {journal} {The European Physical Journal Special
  Topics}\ }\textbf {\bibinfo {volume} {189}},\ \bibinfo {pages} {103}
  (\bibinfo {year} {2010})}\BibitemShut {NoStop}%
\bibitem [{\citenamefont {Liu}\ and\ \citenamefont {et~al.}(2011)}]{Liu2011}%
  \BibitemOpen
  \bibfield  {author} {\bibinfo {author} {\bibfnamefont {C.}~\bibnamefont
  {Liu}}\ and\ \bibinfo {author} {\bibnamefont {et~al.}},\ }\href@noop {}
  {\bibfield  {journal} {\bibinfo  {journal} {Science}\ }\textbf {\bibinfo
  {volume} {334}},\ \bibinfo {pages} {238} (\bibinfo {year}
  {2011})}\BibitemShut {NoStop}%
\bibitem [{\citenamefont {Fu}\ \emph {et~al.}(2012)\citenamefont {Fu},
  \citenamefont {Tang}, \citenamefont {Liu}, \citenamefont {Huang},
  \citenamefont {Hwa},\ and\ \citenamefont {Lenz}}]{Fu2012}%
  \BibitemOpen
  \bibfield  {author} {\bibinfo {author} {\bibfnamefont {X.}~\bibnamefont
  {Fu}}, \bibinfo {author} {\bibfnamefont {L.-H.}\ \bibnamefont {Tang}},
  \bibinfo {author} {\bibfnamefont {C.}~\bibnamefont {Liu}}, \bibinfo {author}
  {\bibfnamefont {J.-D.}\ \bibnamefont {Huang}}, \bibinfo {author}
  {\bibfnamefont {T.}~\bibnamefont {Hwa}}, \ and\ \bibinfo {author}
  {\bibfnamefont {P.}~\bibnamefont {Lenz}},\ }\href@noop {} {\bibfield
  {journal} {\bibinfo  {journal} {Phys. Rev. Lett.}\ }\textbf {\bibinfo
  {volume} {108}},\ \bibinfo {pages} {198102} (\bibinfo {year}
  {2012})}\BibitemShut {NoStop}%
\bibitem [{\citenamefont {Cates}\ \emph {et~al.}(2010)\citenamefont {Cates},
  \citenamefont {Marenduzzo}, \citenamefont {Pagonabarraga},\ and\
  \citenamefont {Tailleur}}]{CatesPNAS2010}%
  \BibitemOpen
  \bibfield  {author} {\bibinfo {author} {\bibfnamefont {M.}~\bibnamefont
  {Cates}}, \bibinfo {author} {\bibfnamefont {D.}~\bibnamefont {Marenduzzo}},
  \bibinfo {author} {\bibfnamefont {I.}~\bibnamefont {Pagonabarraga}}, \ and\
  \bibinfo {author} {\bibfnamefont {J.}~\bibnamefont {Tailleur}},\ }\href@noop
  {} {\bibfield  {journal} {\bibinfo  {journal} {Proc. Nat. Acad. Sci. USA}\
  }\textbf {\bibinfo {volume} {107}},\ \bibinfo {pages} {11715} (\bibinfo
  {year} {2010})}\BibitemShut {NoStop}%
\bibitem [{\citenamefont {B{\"a}uerle}\ \emph {et~al.}(2018)\citenamefont
  {B{\"a}uerle}, \citenamefont {Fischer}, \citenamefont {Speck},\ and\
  \citenamefont {Bechinger}}]{bauerle2018self}%
  \BibitemOpen
  \bibfield  {author} {\bibinfo {author} {\bibfnamefont {T.}~\bibnamefont
  {B{\"a}uerle}}, \bibinfo {author} {\bibfnamefont {A.}~\bibnamefont
  {Fischer}}, \bibinfo {author} {\bibfnamefont {T.}~\bibnamefont {Speck}}, \
  and\ \bibinfo {author} {\bibfnamefont {C.}~\bibnamefont {Bechinger}},\
  }\href@noop {} {\bibfield  {journal} {\bibinfo  {journal} {Nature
  communications}\ }\textbf {\bibinfo {volume} {9}},\ \bibinfo {pages} {1}
  (\bibinfo {year} {2018})}\BibitemShut {NoStop}%
\bibitem [{\citenamefont {Velasco}\ \emph {et~al.}(2018)\citenamefont
  {Velasco}, \citenamefont {Abkenar}, \citenamefont {Gompper},\ and\
  \citenamefont {Auth}}]{velasco2018collective}%
  \BibitemOpen
  \bibfield  {author} {\bibinfo {author} {\bibfnamefont {C.~A.}\ \bibnamefont
  {Velasco}}, \bibinfo {author} {\bibfnamefont {M.}~\bibnamefont {Abkenar}},
  \bibinfo {author} {\bibfnamefont {G.}~\bibnamefont {Gompper}}, \ and\
  \bibinfo {author} {\bibfnamefont {T.}~\bibnamefont {Auth}},\ }\href@noop {}
  {\bibfield  {journal} {\bibinfo  {journal} {Physical Review E}\ }\textbf
  {\bibinfo {volume} {98}},\ \bibinfo {pages} {022605} (\bibinfo {year}
  {2018})}\BibitemShut {NoStop}%
\bibitem [{\citenamefont {Fischer}\ \emph {et~al.}(2020)\citenamefont
  {Fischer}, \citenamefont {Schmid},\ and\ \citenamefont
  {Speck}}]{fischer2020quorum}%
  \BibitemOpen
  \bibfield  {author} {\bibinfo {author} {\bibfnamefont {A.}~\bibnamefont
  {Fischer}}, \bibinfo {author} {\bibfnamefont {F.}~\bibnamefont {Schmid}}, \
  and\ \bibinfo {author} {\bibfnamefont {T.}~\bibnamefont {Speck}},\
  }\href@noop {} {\bibfield  {journal} {\bibinfo  {journal} {Physical Review
  E}\ }\textbf {\bibinfo {volume} {101}},\ \bibinfo {pages} {012601} (\bibinfo
  {year} {2020})}\BibitemShut {NoStop}%
\bibitem [{\citenamefont {Rein}\ \emph {et~al.}(2016)\citenamefont {Rein},
  \citenamefont {Hein{\ss}}, \citenamefont {Schmid},\ and\ \citenamefont
  {Speck}}]{rein2016collective}%
  \BibitemOpen
  \bibfield  {author} {\bibinfo {author} {\bibfnamefont {M.}~\bibnamefont
  {Rein}}, \bibinfo {author} {\bibfnamefont {N.}~\bibnamefont {Hein{\ss}}},
  \bibinfo {author} {\bibfnamefont {F.}~\bibnamefont {Schmid}}, \ and\ \bibinfo
  {author} {\bibfnamefont {T.}~\bibnamefont {Speck}},\ }\href@noop {}
  {\bibfield  {journal} {\bibinfo  {journal} {Physical review letters}\
  }\textbf {\bibinfo {volume} {116}},\ \bibinfo {pages} {058102} (\bibinfo
  {year} {2016})}\BibitemShut {NoStop}%
\bibitem [{\citenamefont {Miller}\ and\ \citenamefont
  {Bassler}(2001)}]{miller2001quorum}%
  \BibitemOpen
  \bibfield  {author} {\bibinfo {author} {\bibfnamefont {M.~B.}\ \bibnamefont
  {Miller}}\ and\ \bibinfo {author} {\bibfnamefont {B.~L.}\ \bibnamefont
  {Bassler}},\ }\href@noop {} {\bibfield  {journal} {\bibinfo  {journal}
  {Annual Reviews in Microbiology}\ }\textbf {\bibinfo {volume} {55}},\
  \bibinfo {pages} {165} (\bibinfo {year} {2001})}\BibitemShut {NoStop}%
\bibitem [{\citenamefont {Katz}\ \emph {et~al.}(2011)\citenamefont {Katz},
  \citenamefont {Tunstrom}, \citenamefont {Ioannou}, \citenamefont {Huepe},\
  and\ \citenamefont {Couzin}}]{Katz2011}%
  \BibitemOpen
  \bibfield  {author} {\bibinfo {author} {\bibfnamefont {Y.}~\bibnamefont
  {Katz}}, \bibinfo {author} {\bibfnamefont {K.}~\bibnamefont {Tunstrom}},
  \bibinfo {author} {\bibfnamefont {C.}~\bibnamefont {Ioannou}}, \bibinfo
  {author} {\bibfnamefont {C.}~\bibnamefont {Huepe}}, \ and\ \bibinfo {author}
  {\bibfnamefont {I.}~\bibnamefont {Couzin}},\ }\href@noop {} {\bibfield
  {journal} {\bibinfo  {journal} {Proc. Nat. Acad. Sci. USA}\ }\textbf
  {\bibinfo {volume} {46}},\ \bibinfo {pages} {18720} (\bibinfo {year}
  {2011})}\BibitemShut {NoStop}%
\bibitem [{\citenamefont {Mishra}\ \emph {et~al.}(2012)\citenamefont {Mishra},
  \citenamefont {Tunstrom}, \citenamefont {Couzin},\ and\ \citenamefont
  {Huepe}}]{Mishra2012}%
  \BibitemOpen
  \bibfield  {author} {\bibinfo {author} {\bibfnamefont {S.}~\bibnamefont
  {Mishra}}, \bibinfo {author} {\bibfnamefont {K.}~\bibnamefont {Tunstrom}},
  \bibinfo {author} {\bibfnamefont {I.}~\bibnamefont {Couzin}}, \ and\ \bibinfo
  {author} {\bibfnamefont {C.}~\bibnamefont {Huepe}},\ }\href@noop {}
  {\bibfield  {journal} {\bibinfo  {journal} {Phys. Rev. E}\ }\textbf {\bibinfo
  {volume} {86}},\ \bibinfo {pages} {011901} (\bibinfo {year}
  {2012})}\BibitemShut {NoStop}%
\bibitem [{\citenamefont {Licata}\ \emph {et~al.}(2016)\citenamefont {Licata},
  \citenamefont {Mohari}, \citenamefont {Fuqua},\ and\ \citenamefont
  {Setaysehgar}}]{Licata2016}%
  \BibitemOpen
  \bibfield  {author} {\bibinfo {author} {\bibfnamefont {N.}~\bibnamefont
  {Licata}}, \bibinfo {author} {\bibfnamefont {B.}~\bibnamefont {Mohari}},
  \bibinfo {author} {\bibfnamefont {C.}~\bibnamefont {Fuqua}}, \ and\ \bibinfo
  {author} {\bibfnamefont {S.}~\bibnamefont {Setaysehgar}},\ }\href@noop {}
  {\bibfield  {journal} {\bibinfo  {journal} {Biophys. J.}\ }\textbf {\bibinfo
  {volume} {110}},\ \bibinfo {pages} {247} (\bibinfo {year}
  {2016})}\BibitemShut {NoStop}%
\bibitem [{\citenamefont {Bhattacharjee}\ \emph {et~al.}(2021)\citenamefont
  {Bhattacharjee}, \citenamefont {Amchin}, \citenamefont {Ott}, \citenamefont
  {Kratz},\ and\ \citenamefont {Datta}}]{bhattacharjee2021chemotactic}%
  \BibitemOpen
  \bibfield  {author} {\bibinfo {author} {\bibfnamefont {T.}~\bibnamefont
  {Bhattacharjee}}, \bibinfo {author} {\bibfnamefont {D.~B.}\ \bibnamefont
  {Amchin}}, \bibinfo {author} {\bibfnamefont {J.~A.}\ \bibnamefont {Ott}},
  \bibinfo {author} {\bibfnamefont {F.}~\bibnamefont {Kratz}}, \ and\ \bibinfo
  {author} {\bibfnamefont {S.~S.}\ \bibnamefont {Datta}},\ }\href@noop {}
  {\bibfield  {journal} {\bibinfo  {journal} {Biophysical Journal}\ } (\bibinfo
  {year} {2021})}\BibitemShut {NoStop}%
\bibitem [{\citenamefont {Volpe}\ and\ \citenamefont
  {Volpe}(2017)}]{Volpe2017}%
  \BibitemOpen
  \bibfield  {author} {\bibinfo {author} {\bibfnamefont {G.}~\bibnamefont
  {Volpe}}\ and\ \bibinfo {author} {\bibfnamefont {G.}~\bibnamefont {Volpe}},\
  }\href@noop {} {\bibfield  {journal} {\bibinfo  {journal} {Proc. Nat. Acad.
  Sci. USA}\ }\textbf {\bibinfo {volume} {114}},\ \bibinfo {pages} {11350}
  (\bibinfo {year} {2017})}\BibitemShut {NoStop}%
\bibitem [{\citenamefont {Berke}\ \emph {et~al.}(2008)\citenamefont {Berke},
  \citenamefont {Turner}, \citenamefont {Berg},\ and\ \citenamefont
  {Lauga}}]{berke2008hydrodynamic}%
  \BibitemOpen
  \bibfield  {author} {\bibinfo {author} {\bibfnamefont {A.~P.}\ \bibnamefont
  {Berke}}, \bibinfo {author} {\bibfnamefont {L.}~\bibnamefont {Turner}},
  \bibinfo {author} {\bibfnamefont {H.~C.}\ \bibnamefont {Berg}}, \ and\
  \bibinfo {author} {\bibfnamefont {E.}~\bibnamefont {Lauga}},\ }\href@noop {}
  {\bibfield  {journal} {\bibinfo  {journal} {Physical Review Letters}\
  }\textbf {\bibinfo {volume} {101}},\ \bibinfo {pages} {038102} (\bibinfo
  {year} {2008})}\BibitemShut {NoStop}%
\bibitem [{\citenamefont {Takagi}\ \emph {et~al.}(2014)\citenamefont {Takagi},
  \citenamefont {Palacci}, \citenamefont {Braunschweig}, \citenamefont
  {Shelley},\ and\ \citenamefont {Zhang}}]{takagi2014hydrodynamic}%
  \BibitemOpen
  \bibfield  {author} {\bibinfo {author} {\bibfnamefont {D.}~\bibnamefont
  {Takagi}}, \bibinfo {author} {\bibfnamefont {J.}~\bibnamefont {Palacci}},
  \bibinfo {author} {\bibfnamefont {A.~B.}\ \bibnamefont {Braunschweig}},
  \bibinfo {author} {\bibfnamefont {M.~J.}\ \bibnamefont {Shelley}}, \ and\
  \bibinfo {author} {\bibfnamefont {J.}~\bibnamefont {Zhang}},\ }\href@noop {}
  {\bibfield  {journal} {\bibinfo  {journal} {Soft Matter}\ }\textbf {\bibinfo
  {volume} {10}},\ \bibinfo {pages} {1784} (\bibinfo {year}
  {2014})}\BibitemShut {NoStop}%
\bibitem [{\citenamefont {Sipos}\ \emph {et~al.}(2015)\citenamefont {Sipos},
  \citenamefont {Nagy}, \citenamefont {Di~Leonardo},\ and\ \citenamefont
  {Galajda}}]{sipos2015hydrodynamic}%
  \BibitemOpen
  \bibfield  {author} {\bibinfo {author} {\bibfnamefont {O.}~\bibnamefont
  {Sipos}}, \bibinfo {author} {\bibfnamefont {K.}~\bibnamefont {Nagy}},
  \bibinfo {author} {\bibfnamefont {R.}~\bibnamefont {Di~Leonardo}}, \ and\
  \bibinfo {author} {\bibfnamefont {P.}~\bibnamefont {Galajda}},\ }\href@noop
  {} {\bibfield  {journal} {\bibinfo  {journal} {Physical review letters}\
  }\textbf {\bibinfo {volume} {114}},\ \bibinfo {pages} {258104} (\bibinfo
  {year} {2015})}\BibitemShut {NoStop}%
\bibitem [{\citenamefont {Drescher}\ \emph {et~al.}(2011)\citenamefont
  {Drescher}, \citenamefont {Dunkel}, \citenamefont {Cisneros}, \citenamefont
  {Ganguly},\ and\ \citenamefont {Goldstein}}]{drescher2011fluid}%
  \BibitemOpen
  \bibfield  {author} {\bibinfo {author} {\bibfnamefont {K.}~\bibnamefont
  {Drescher}}, \bibinfo {author} {\bibfnamefont {J.}~\bibnamefont {Dunkel}},
  \bibinfo {author} {\bibfnamefont {L.~H.}\ \bibnamefont {Cisneros}}, \bibinfo
  {author} {\bibfnamefont {S.}~\bibnamefont {Ganguly}}, \ and\ \bibinfo
  {author} {\bibfnamefont {R.~E.}\ \bibnamefont {Goldstein}},\ }\href@noop {}
  {\bibfield  {journal} {\bibinfo  {journal} {Proceedings of the National
  Academy of Sciences}\ }\textbf {\bibinfo {volume} {108}},\ \bibinfo {pages}
  {10940} (\bibinfo {year} {2011})}\BibitemShut {NoStop}%
\bibitem [{\citenamefont {Molaei}\ and\ \citenamefont
  {Sheng}(2016)}]{molaei2016succeed}%
  \BibitemOpen
  \bibfield  {author} {\bibinfo {author} {\bibfnamefont {M.}~\bibnamefont
  {Molaei}}\ and\ \bibinfo {author} {\bibfnamefont {J.}~\bibnamefont {Sheng}},\
  }\href@noop {} {\bibfield  {journal} {\bibinfo  {journal} {Scientific
  reports}\ }\textbf {\bibinfo {volume} {6}},\ \bibinfo {pages} {1} (\bibinfo
  {year} {2016})}\BibitemShut {NoStop}%
\bibitem [{\citenamefont {Fahrner}\ \emph {et~al.}(2003)\citenamefont
  {Fahrner}, \citenamefont {Ryu},\ and\ \citenamefont
  {Berg}}]{fahrner2003bacterial}%
  \BibitemOpen
  \bibfield  {author} {\bibinfo {author} {\bibfnamefont {K.~A.}\ \bibnamefont
  {Fahrner}}, \bibinfo {author} {\bibfnamefont {W.~S.}\ \bibnamefont {Ryu}}, \
  and\ \bibinfo {author} {\bibfnamefont {H.~C.}\ \bibnamefont {Berg}},\
  }\href@noop {} {\bibfield  {journal} {\bibinfo  {journal} {Nature}\ }\textbf
  {\bibinfo {volume} {423}},\ \bibinfo {pages} {938} (\bibinfo {year}
  {2003})}\BibitemShut {NoStop}%
\bibitem [{\citenamefont {Tipping}\ \emph {et~al.}(2013)\citenamefont
  {Tipping}, \citenamefont {Delalez}, \citenamefont {Lim}, \citenamefont
  {Berry},\ and\ \citenamefont {Armitage}}]{tipping2013load}%
  \BibitemOpen
  \bibfield  {author} {\bibinfo {author} {\bibfnamefont {M.~J.}\ \bibnamefont
  {Tipping}}, \bibinfo {author} {\bibfnamefont {N.~J.}\ \bibnamefont
  {Delalez}}, \bibinfo {author} {\bibfnamefont {R.}~\bibnamefont {Lim}},
  \bibinfo {author} {\bibfnamefont {R.~M.}\ \bibnamefont {Berry}}, \ and\
  \bibinfo {author} {\bibfnamefont {J.~P.}\ \bibnamefont {Armitage}},\
  }\href@noop {} {\bibfield  {journal} {\bibinfo  {journal} {MBio}\ }\textbf
  {\bibinfo {volume} {4}},\ \bibinfo {pages} {e00551} (\bibinfo {year}
  {2013})}\BibitemShut {NoStop}%
\bibitem [{\citenamefont {Wadhwa}\ \emph {et~al.}(2019)\citenamefont {Wadhwa},
  \citenamefont {Phillips},\ and\ \citenamefont {Berg}}]{wadhwa2019torque}%
  \BibitemOpen
  \bibfield  {author} {\bibinfo {author} {\bibfnamefont {N.}~\bibnamefont
  {Wadhwa}}, \bibinfo {author} {\bibfnamefont {R.}~\bibnamefont {Phillips}}, \
  and\ \bibinfo {author} {\bibfnamefont {H.~C.}\ \bibnamefont {Berg}},\
  }\href@noop {} {\bibfield  {journal} {\bibinfo  {journal} {Proceedings of the
  National Academy of Sciences}\ }\textbf {\bibinfo {volume} {116}},\ \bibinfo
  {pages} {11764} (\bibinfo {year} {2019})}\BibitemShut {NoStop}%
\bibitem [{\citenamefont {Anderson}\ \emph {et~al.}(2020)\citenamefont
  {Anderson}, \citenamefont {Glaser},\ and\ \citenamefont
  {Glotzer}}]{HOOMD2020}%
  \BibitemOpen
  \bibfield  {author} {\bibinfo {author} {\bibfnamefont {J.~A.}\ \bibnamefont
  {Anderson}}, \bibinfo {author} {\bibfnamefont {J.}~\bibnamefont {Glaser}}, \
  and\ \bibinfo {author} {\bibfnamefont {S.~C.}\ \bibnamefont {Glotzer}},\
  }\href {\doibase https://doi.org/10.1016/j.commatsci.2019.109363} {\bibfield
  {journal} {\bibinfo  {journal} {Computational Materials Science}\ }\textbf
  {\bibinfo {volume} {173}},\ \bibinfo {pages} {109363} (\bibinfo {year}
  {2020})}\BibitemShut {NoStop}%
\bibitem [{\citenamefont {Ramasubramani}\ \emph {et~al.}(2020)\citenamefont
  {Ramasubramani}, \citenamefont {Dice}, \citenamefont {Harper}, \citenamefont
  {Spellings}, \citenamefont {Anderson},\ and\ \citenamefont
  {Glotzer}}]{freud2020}%
  \BibitemOpen
  \bibfield  {author} {\bibinfo {author} {\bibfnamefont {V.}~\bibnamefont
  {Ramasubramani}}, \bibinfo {author} {\bibfnamefont {B.~D.}\ \bibnamefont
  {Dice}}, \bibinfo {author} {\bibfnamefont {E.~S.}\ \bibnamefont {Harper}},
  \bibinfo {author} {\bibfnamefont {M.~P.}\ \bibnamefont {Spellings}}, \bibinfo
  {author} {\bibfnamefont {J.~A.}\ \bibnamefont {Anderson}}, \ and\ \bibinfo
  {author} {\bibfnamefont {S.~C.}\ \bibnamefont {Glotzer}},\ }\href {\doibase
  https://doi.org/10.1016/j.cpc.2020.107275} {\bibfield  {journal} {\bibinfo
  {journal} {Computer Physics Communications}\ }\textbf {\bibinfo {volume}
  {254}},\ \bibinfo {pages} {107275} (\bibinfo {year} {2020})}\BibitemShut
  {NoStop}%
\bibitem [{\citenamefont {Torquato}\ and\ \citenamefont
  {Lu}(1993)}]{Torquato1993}%
  \BibitemOpen
  \bibfield  {author} {\bibinfo {author} {\bibfnamefont {S.}~\bibnamefont
  {Torquato}}\ and\ \bibinfo {author} {\bibfnamefont {B.}~\bibnamefont {Lu}},\
  }\href@noop {} {\bibfield  {journal} {\bibinfo  {journal} {Phys. Rev. E}\
  }\textbf {\bibinfo {volume} {47}},\ \bibinfo {pages} {2950} (\bibinfo {year}
  {1993})}\BibitemShut {NoStop}%
\bibitem [{sup()}]{suppmat}%
  \BibitemOpen
  \href@noop {} {}\bibinfo {note} {Supplementary Materials: Movies and extra
  plots regarding the trimer's displacements and glassy dynamics.}\BibitemShut
  {Stop}%
\bibitem [{\citenamefont {Schnyder}\ \emph {et~al.}(2015)\citenamefont
  {Schnyder}, \citenamefont {Spanner}, \citenamefont {H{\"o}fling},
  \citenamefont {Franosch},\ and\ \citenamefont
  {Horbach}}]{schnyder2015rounding}%
  \BibitemOpen
  \bibfield  {author} {\bibinfo {author} {\bibfnamefont {S.~K.}\ \bibnamefont
  {Schnyder}}, \bibinfo {author} {\bibfnamefont {M.}~\bibnamefont {Spanner}},
  \bibinfo {author} {\bibfnamefont {F.}~\bibnamefont {H{\"o}fling}}, \bibinfo
  {author} {\bibfnamefont {T.}~\bibnamefont {Franosch}}, \ and\ \bibinfo
  {author} {\bibfnamefont {J.}~\bibnamefont {Horbach}},\ }\href@noop {}
  {\bibfield  {journal} {\bibinfo  {journal} {Soft Matter}\ }\textbf {\bibinfo
  {volume} {11}},\ \bibinfo {pages} {701} (\bibinfo {year} {2015})}\BibitemShut
  {NoStop}%
\end{thebibliography}%


\begin{thebibliography}{3}%
\makeatletter
\providecommand \@ifxundefined [1]{%
 \@ifx{#1\undefined}
}%
\providecommand \@ifnum [1]{%
 \ifnum #1\expandafter \@firstoftwo
 \else \expandafter \@secondoftwo
 \fi
}%
\providecommand \@ifx [1]{%
 \ifx #1\expandafter \@firstoftwo
 \else \expandafter \@secondoftwo
 \fi
}%
\providecommand \natexlab [1]{#1}%
\providecommand \enquote  [1]{``#1''}%
\providecommand \bibnamefont  [1]{#1}%
\providecommand \bibfnamefont [1]{#1}%
\providecommand \citenamefont [1]{#1}%
\providecommand \href@noop [0]{\@secondoftwo}%
\providecommand \href [0]{\begingroup \@sanitize@url \@href}%
\providecommand \@href[1]{\@@startlink{#1}\@@href}%
\providecommand \@@href[1]{\endgroup#1\@@endlink}%
\providecommand \@sanitize@url [0]{\catcode `\\12\catcode `\$12\catcode
  `\&12\catcode `\#12\catcode `\^12\catcode `\_12\catcode `\%12\relax}%
\providecommand \@@startlink[1]{}%
\providecommand \@@endlink[0]{}%
\providecommand \url  [0]{\begingroup\@sanitize@url \@url }%
\providecommand \@url [1]{\endgroup\@href {#1}{\urlprefix }}%
\providecommand \urlprefix  [0]{URL }%
\providecommand \Eprint [0]{\href }%
\providecommand \doibase [0]{https://doi.org/}%
\providecommand \selectlanguage [0]{\@gobble}%
\providecommand \bibinfo  [0]{\@secondoftwo}%
\providecommand \bibfield  [0]{\@secondoftwo}%
\providecommand \translation [1]{[#1]}%
\providecommand \BibitemOpen [0]{}%
\providecommand \bibitemStop [0]{}%
\providecommand \bibitemNoStop [0]{.\EOS\space}%
\providecommand \EOS [0]{\spacefactor3000\relax}%
\providecommand \BibitemShut  [1]{\csname bibitem#1\endcsname}%
\let\auto@bib@innerbib\@empty
\bibitem [{\citenamefont {Stukowski}(2010)}]{ovito}%
  \BibitemOpen
  \bibfield  {author} {\bibinfo {author} {\bibfnamefont {A.}~\bibnamefont
  {Stukowski}},\ }\bibfield  {title} {\bibinfo {title} {{Visualization and
  analysis of atomistic simulation data with OVITO-the Open Visualization
  Tool}},\ }\bibfield  {journal} {\bibinfo  {journal} {{MODELLING AND
  SIMULATION IN MATERIALS SCIENCE AND ENGINEERING}}\ }\textbf {\bibinfo
  {volume} {{18}}},\ \href {https://doi.org/{10.1088/0965-0393/18/1/015012}}
  {{10.1088/0965-0393/18/1/015012}} (\bibinfo {year} {{2010}})\BibitemShut
  {NoStop}%
\bibitem [{\citenamefont {Kurzthaler}\ \emph {et~al.}(2021)\citenamefont
  {Kurzthaler}, \citenamefont {Mandal}, \citenamefont {Bhattacharjee},
  \citenamefont {L{\"o}wen}, \citenamefont {Datta},\ and\ \citenamefont
  {Stone}}]{kurzthaler2021geometric}%
  \BibitemOpen
  \bibfield  {author} {\bibinfo {author} {\bibfnamefont {C.}~\bibnamefont
  {Kurzthaler}}, \bibinfo {author} {\bibfnamefont {S.}~\bibnamefont {Mandal}},
  \bibinfo {author} {\bibfnamefont {T.}~\bibnamefont {Bhattacharjee}}, \bibinfo
  {author} {\bibfnamefont {H.}~\bibnamefont {L{\"o}wen}}, \bibinfo {author}
  {\bibfnamefont {S.~S.}\ \bibnamefont {Datta}},\ and\ \bibinfo {author}
  {\bibfnamefont {H.~A.}\ \bibnamefont {Stone}},\ }\bibfield  {title} {\bibinfo
  {title} {A geometric criterion for the optimal spreading of active polymers
  in porous media},\ }\href {https://doi.org/10.1038/s41467-021-26942-0}
  {\bibfield  {journal} {\bibinfo  {journal} {Nature Communications}\ }\textbf
  {\bibinfo {volume} {12}},\ \bibinfo {pages} {7088} (\bibinfo {year}
  {2021})}\BibitemShut {NoStop}%
\bibitem [{\citenamefont {Bhattacharjee}\ and\ \citenamefont
  {Datta}(2019)}]{datta2019}%
  \BibitemOpen
  \bibfield  {author} {\bibinfo {author} {\bibfnamefont {T.}~\bibnamefont
  {Bhattacharjee}}\ and\ \bibinfo {author} {\bibfnamefont {S.}~\bibnamefont
  {Datta}},\ }\bibfield  {title} {\bibinfo {title} {Bacterial hopping and
  trapping in porous media},\ }\href@noop {} {\bibfield  {journal} {\bibinfo
  {journal} {Nature Communications}\ }\textbf {\bibinfo {volume} {10}},\
  \bibinfo {pages} {1} (\bibinfo {year} {2019})}\BibitemShut {NoStop}%
\end{thebibliography}%
\end{document}


\title{Dynamics of bacteria scanning a porous environment:\\Supplementary Materials}
\author{Ehsan Irani}
\affiliation{Max Delbrück Center for Molecular Medicine in the Helmholtz Association(MDC), The Berlin Institute for Medical Systems Biology (BIMSB), Berlin, Germany}
\author{Zahra Mokhtari}
\affiliation{Department of Mathematics and Computer Science, Freie Universit\"at Berlin}
\author{Annette Zippelius}
\affiliation{Institut f\"ur Theoretische Physik, Universit\"at G\"ottingen}
\date{}
\maketitle

\section{Movies}
\label{sec:orgb58be17}
The  movies  illustrate  the  qualitative  difference  between the transport of trimers with density-dependent and uniform reorientations in porous media. To study the single agent behavior in porous media we have turned off the interaction between different trimers.  In following movies \(\phi_{\rm{o}}\) and \(G\)  vary while \(\tau_{\rm{re}}=5.0\):
\begin{itemize}
\item Sigmoidal G: Here \(G\) is a sigmoidal function of the local density, as given by Eq. 3 in the main text. Active trimers traverse large distances almost ballistically and reorient almost only at dead ends, where two obstacles form a wedge-shaped trap.
\begin{itemize}
\item \texttt{phi\_obs0.4-sigmoid.webm}
\item \texttt{phi\_obs0.6-sigmoid.webm}
\end{itemize}
\item Constant G: Here \(G\) is a constant, giving rise to uniform reorientations. The trimers do not sense the environment and reorient with equal probability in the void space or in local cages.
\begin{itemize}
\item \texttt{phi\_obs0.4-uniform.webm}
\item \texttt{phi\_obs0.6-uniform.webm}
\end{itemize}
\end{itemize}

For trimers with sigmoidal G, the scanning area around the trimer is displayed by a disk colored by the value of \(\phi_{\rm{local}}\): Blue for small values, white for intermediate values and red for high values. OVITO \cite{ovito} is used to visualize simulations and make movies.

\section{Reverse dynamics vs. random reorientation}

\begin{figure}[!htb]
\centering
\includegraphics[width=0.48\textwidth]{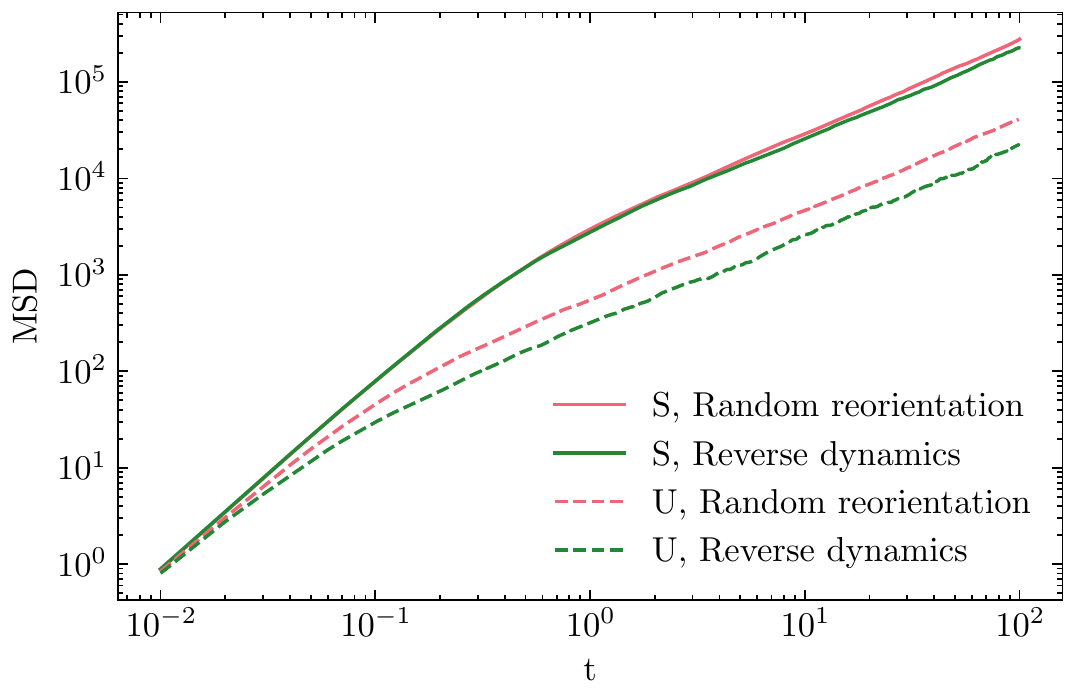}
\caption{\label{fig:reverse} MSD for a sigmoidal function $G$ (full line) and constant $G$ (dashed line), comparing reorientation in the form of reversals to randomly chosen angles.}
\end{figure}
It has been suggested recently~\cite{kurzthaler2021geometric} that
reversing the velocity of the active particle is an efficient means to
accelerate the dynamics. For the sigmoidal $G$ random reorientation and run-reverse dynamics can hardly be distinguished. In Fig.~\ref{fig:reverse} we
compare random reorientation and run-reverse dynamics for both
constant $G$ and quorum sensing. The acceleration of the dynamics due to density-sensing is even
stronger for run-reverse dynamics (dashed and full green line) than
for uniform reorientation (dashed and full red line).

\section{Reorientation time: Comparison with the experimental data}

Given our simulation of a two-dimensional system, we can only aim for
a qualitative comparison to the three-dimensional data of the
experiment~\cite{datta2019}.
In the experiment the bacteria are trapped for times
$t_{\rm{tr}}$, which are broadly distributed over a range of trapping
times $\SI{0.4}{s}...\SI{40}{s}$ (Fig.4b of ref.~\cite{datta2019} ).
The relevant quantity to compare to experiment is not $\tau_{\rm{re}}$
but the waiting time: the time between two successive reorientation
events. The waiting times of our simulation are also broadly
distributed as shown in Fig.3a of the main text. To compare absolute
numbers, we estimate the experimental value for $t_a\sim\SI{0.1}{s}$
(running speed is about $\SI{20}{\micro m \per s}$ and
$2R_t\sim\SI{2}{\micro m}$).  Taking $t_{\rm{tr}}=\SI{4}{s}$ as a
``typical'' value, the ratio is given by $t_{\rm{tr}}/t_a\sim
40$. This estimate agrees with our data for the trapping times at high
densities (see Fig.5c in the main text) and is furthermore comparable
to our theoretical values for the mean waiting time shown in the inset
of Fig.3b. However we want to stress again that the comparison is at
best qualitative, given that we simulate a two-dimensional system.


\section{Glassy dynamics}
\label{sec:org37a82c8}
\begin{figure}[!htb]
\centering
\includegraphics[width=0.48\textwidth]{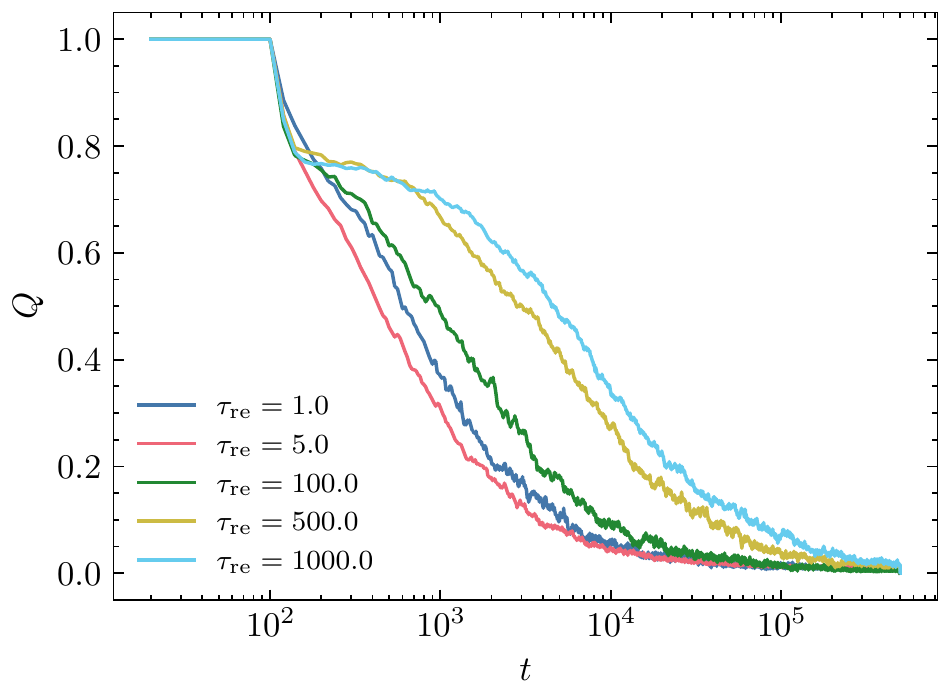}
\caption{\label{fig:orga9fce5a}\(Q_d(t)\) for \(d=100\), \(\phi_{\rm{o}}=0.4\) and several values of \(\tau_{\rm{re}}\).}
\end{figure}

\begin{figure}[!h]
\centering
\includegraphics[width=0.48\textwidth]{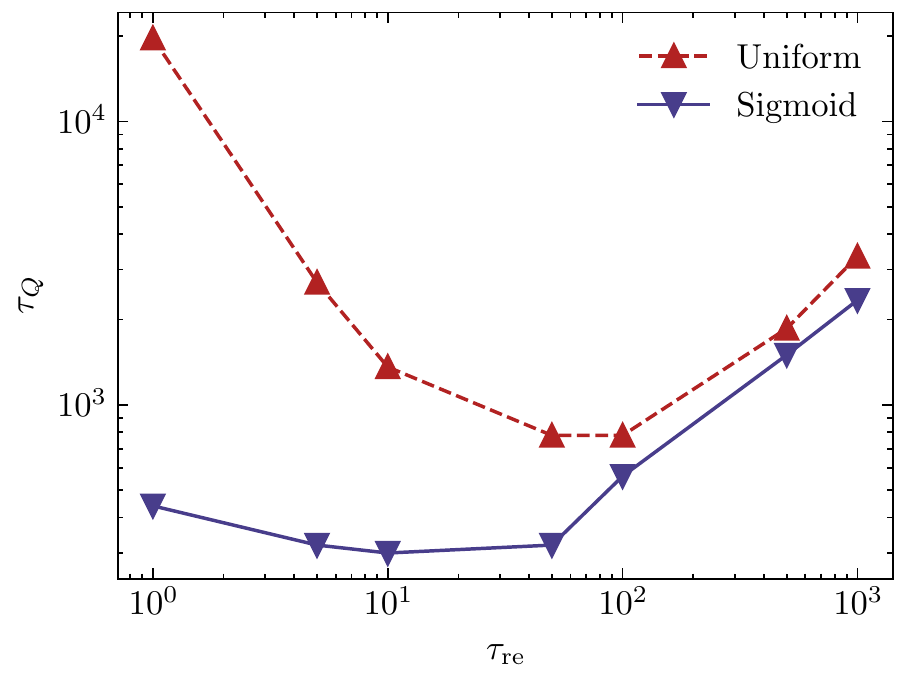}
\caption{\label{fig:org6ddf8e3}\(\tau_Q\) versus \(\tau_{\rm{re}}\) for systems with uniform or sigmoidal G at \(\phi_{\rm{o}}=0.4\).}
\end{figure}

The fraction of localised particles is quantified by \(Q(t;d)\). It is displayed in Fig. \ref{fig:orga9fce5a} for \(d=0.3L\) at \(\phi_{\rm{o}}=0.4\). We observe an intermediate plateau and a second relaxation resembling the glassy dynamics for large reorientation times, \(\tau_{\rm{re}} \geq 500\).
The decay time \(\tau_Q\) is defined as the time at which \(Q(\tau_Q) = 0.6\).  Fig. \ref{fig:org6ddf8e3} exhibits the non-monotonic dependence of \(\tau_Q\) on reorientation time \(\tau_{\rm{re}}\) for both constant and sigmoidal \(G\).

\begin{figure}[!h]
\centering
\includegraphics[width=0.48\textwidth]{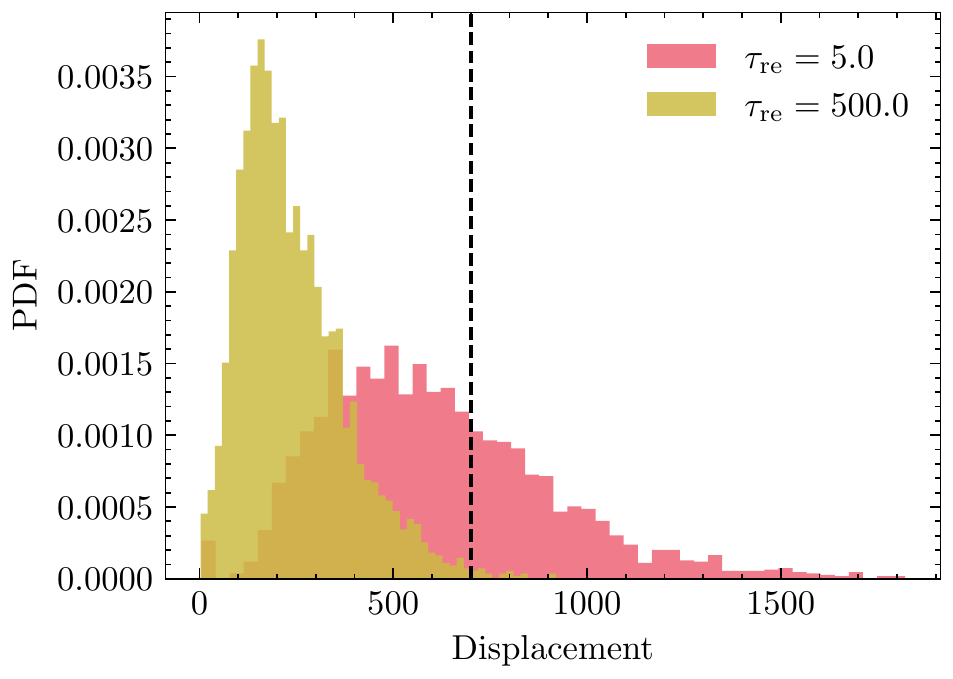}
\caption{\label{fig:org6e2a00d}Distribution of the displacements at \(t=10^4\) at \(\phi_{\text{o}}=0.4\) for two values of \(\tau_{\rm{re}}\), revealing the localisation of the trimers at large \(\tau_{\rm{re}}\). The dashed line indicates the size of the periodic box.}
\end{figure}

Fig. \ref{fig:org6e2a00d} exhibits the distributions of particle displacements at \(t=10^4\). Large reorientation time results in localisation of most of the trimers. However, finite fraction of trimers remain localised with decreasing the reorientation time, even at \(\tau_{\rm{re}}=5.0\) where the highest diffusion coefficient is observed.

\bibliography{Bibliography}